\documentclass[
 reprint,
 superscriptaddress,
 linenumbers
 showpacs,
 nofootinbib,
 amsmath,amssymb,
 aps,
 onecolumn,
 notitlepage
 ]{revtex4-1}

 \usepackage{graphicx}
 \usepackage{dcolumn}
 \usepackage{bm}
 \usepackage{hyperref}
 \usepackage{amsmath}
 \usepackage{amssymb}
 \usepackage{bbold}
 \usepackage{url}
 \usepackage{color}

\def\v#1{{\bf#1}}
\def\be{\begin{equation}}
\def\ee{\end{equation}}
\def\bea{\begin{eqnarray}}
\def\eea{\end{eqnarray}}

\def\<{\langle}
\def\>{\rangle}

\begin{document}

\title{Hidden duality and accidental degeneracy in cycloacene and M\"obius cycloacene}

\author{Emerson Sadurn\'i}
\email{sadurni@ifuap.buap.mx}
\affiliation{Instituto de F\'isica, Benem\'erita Universidad Aut\'onoma de Puebla, Apartado Postal J-48, 72570 Puebla, M\'exico}

\author{Francois Leyvraz}
\affiliation{Instituto de Ciencias F\'isicas, Universidad Nacional Aut\'onoma de M\'exico, 62210 Cuernavaca, Mexico}
\affiliation{Centro Internacional de Ciencias, 62210 Cuernavaca, Mexico}

\author{Thomas Stegmann}
\affiliation{Instituto de Ciencias F\'isicas, Universidad Nacional Aut\'onoma de M\'exico, 62210 Cuernavaca, Mexico}

\author{Thomas H. Seligman}
\email{seligman@icf.unam.mx}
\affiliation{Instituto de Ciencias F\'isicas, Universidad Nacional Aut\'onoma de M\'exico, 62210 Cuernavaca, Mexico}
\affiliation{Centro Internacional de Ciencias, 62210 Cuernavaca, Mexico}

\author{Douglas J. Klein}
\affiliation{Department of Marine Sciences, Texas A\&M University at Galveston, Galveston, TX, 77553-1675, USA}

\date{\today}

\begin{abstract}
  The accidental degeneracy appearing in cycloacenes as triplets and quadruplets is explained with
  the concept of segmentation, introduced here with the aim of describing the effective
  disconnection of $\pi$ orbitals on these organic compounds. For periodic systems with time
  reversal symmetry, the emergent nodal domains are shown to divide the atomic chains into simpler
  carbon structures analog to benzene rings, diallyl chains, anthracene (triacene) chains and
  tetramethyl-naphtalene skeletal forms. The common electronic levels of these segments are
  identified as members of degenerate multiplets of the global system. The peculiar degeneracy of
  M\"obius cycloacene is also explained by segmentation. In the last part, it is shown that the
  multiplicity of energies for cycloacene can be foreseen by studying the continuous limit of the
  tight-binding model; the degeneracy conditions are put in terms of Chebyshev polynomials. The
  results obtained in this work have important consequences on the physics of electronic transport
  in organic wires, together with their artificial realizations.
\end{abstract}



\maketitle


\section{Introduction\label{sec:1}} 

Hidden symmetries and accidental degeneracies play an important role in understanding the paradigms
of nature. The triumph of superintegrability in the explanation of degeneracies has covered not only
our extensive use of harmonic oscillators \cite{ho1, ho3, BS} and Coulomb potentials \cite{cp3, cp4}
in atomic and nuclear physics. Important extensions of these concepts where introduced in five
papers by Moshinsky and various collaborators \cite{BS, Mosh1, Mosh2, Mosh3, Mosh4}, which lead to
further vast generalizations encompassed by Calogero models \cite{calogero, sutherland, khas2000}.

\begin{figure}[b]
  \begin{center}
    \includegraphics[scale=0.3]{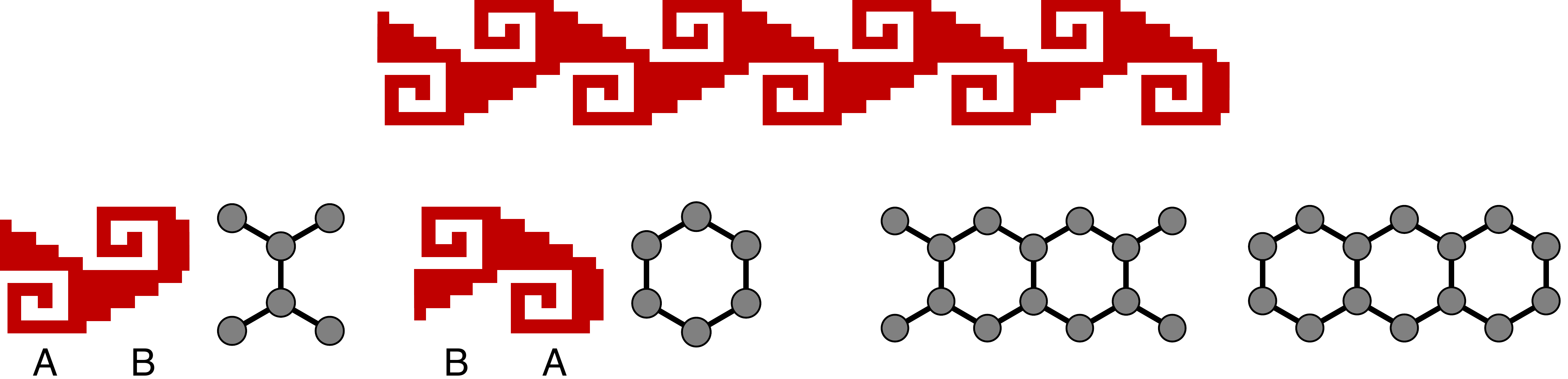}
  \end{center}
  \caption{\label{fig:1} Periodic systems with duality. Upper row: Wall pattern found at the
    pyramids of Mitla (Mexico) show a binary structure. Bottom row, left: Two possible building
    blocks AB and BA, with their organic analogues diallyl and benzene. Bottom row, right: Large
    arrays allow more complex segments, such as tetramethyl-naphtalene and anthracene.}
\end{figure}

\begin{figure}[t]
  \begin{center}
    \includegraphics[scale=0.3]{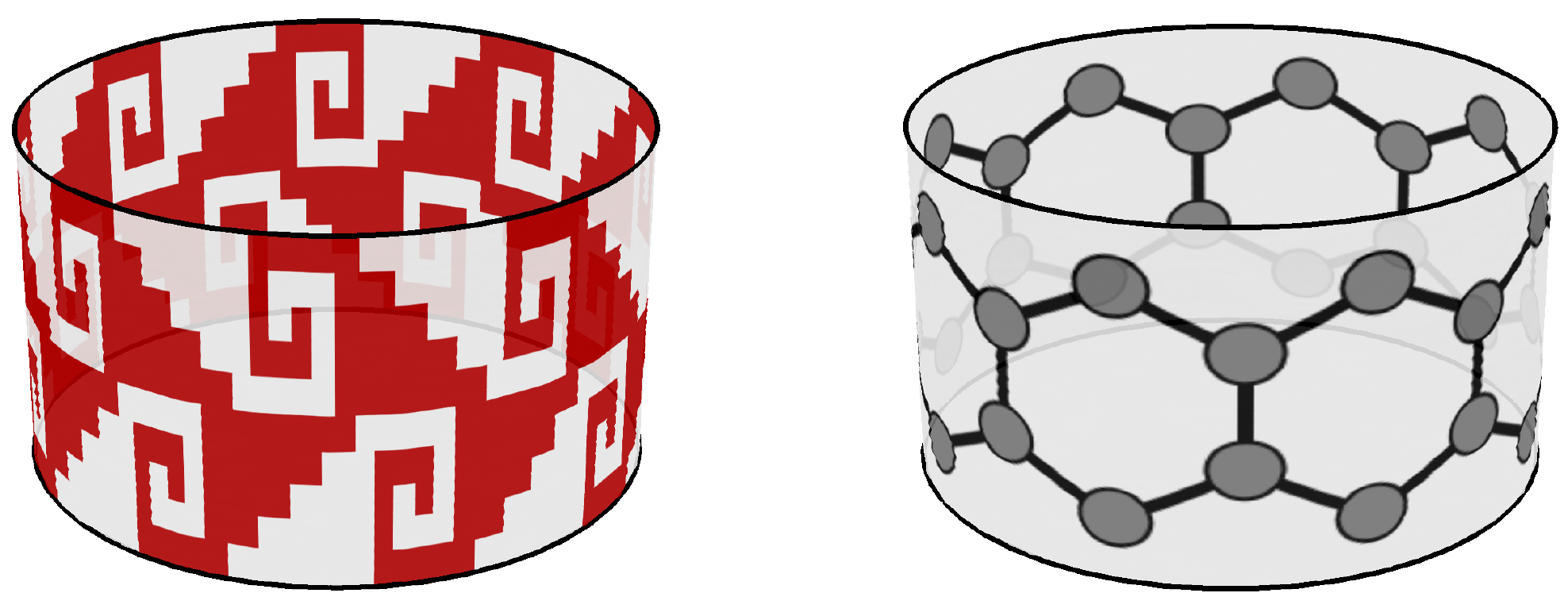}
  \end{center}
  \caption{\label{fig:1.1} Duality wrapped around a cylinder. The right panel shows cyclo-octoacene
    as an example with triplets and quadruplets in the electronic spectrum of carbon $\pi$
    orbitals.}
\end{figure}

However, some accidental degeneracies are far more difficult to explain within the Lie algebraic
context, as it may happen in some judiciously designed systems \cite{sadurni2019, sadurni2016(2)}
where level crossings \cite{kato1966, pechukas1983, gordon2000, steeb1988} and exceptional points
\cite{heiss1990, heiss2004, berry1984, berry1985, ortega2020} can be found. Furthermore, important
examples without external control parameters occur naturally in chemistry, as can be seen in the
electronic spectrum of long organic compounds based on hexagonal structures: cyclopolyacenes
\cite{nakamura2003, houk2001, choi1999, karmakar, salem, kertesz}.  In recent times, these
interesting structures have received considerable attention, spurred by the need for better
electronic transport in large molecules with nanotechnological applications \cite{heeger1993,
  campbell1983, heeger2001, heeger2003, stegmann2017, stegmann2017a, stegmann2020}. Promising
results related to these highly-conducting complexes include organic \cite{tour2001} and inorganic
\cite{mihailovic2009} realizations. Our interest here is to study the degenerate electronic states
associated to skeletal forms of carbon atoms that allow various superpositions of electronic
currents at the same frequency with an important influence on the overall transport properties or
electronic mobility.

We shall see in this paper that a key concept to predict degeneracies in periodic systems is the
so-called duality \cite{sch, coxeter} in tilings. This property is identified here for the first
time in connection with cyclic compounds, stemming from a general formulation of periodic systems
with time reversal invariance. Our approach is further clarified in the context of H\"uckel
(tight-binding) models \cite{hueckel1931(1), hueckel1931(2), hueckel1932, hueckel1933} applied to
the organic compounds of interest. Important precedents in this direction can be found in
\cite{kivelson, bozovic}, and other useful approaches in \cite{garcia, bhattacharya}.

It should be noted, however, that a greater generality related to molecular segmentation was
identified four decades ago. The problem of degeneracy in organic molecules was originally addressed
by G. G. Hall \cite{hall1, hall2, hall3} beyond configurational symmetry. While the application of
graph theory \cite{hall1} to the H\"uckel model of molecules is a rather direct one, some of its
consequences are less trivial: The terms subspectrality \cite{hall2} and cospectrality \cite{hall3}
were coined to describe general graph properties related to the spectrum of molecular fragments, the
latter being contained in the full spectral set of larger molecules. This plays an important role in
our considerations.

The usefulness of our method is such that its application is not restricted to orientable
molecules. The famous case of M\"obius cycloacene also displays accidental triplets as a function of
the molecule's length. The method of segmentation shall be applied to this case as well.

Beyond the segmentation approach, we also study (in the case of cycloacene) the corresponding
infinite system. There we find a degeneracy in the resulting continuous spectrum, which can be
described in simple and general terms, but which does not generate a degeneracy in the corresponding
finite cyclic systems, except in some quite particular cases. We thereby understand the
irregularities of the cycloacene degeneracies as the effect of the boundary conditions on a general
degeneracy occurring in the continuous spectrum for the corresponding infinite system.

\section{Segmentation of periodic systems \label{sec:2}}

Infinite periodic systems and finite systems wrapped around a torus -- as shown in fig.~\ref{fig:1}
and~\ref{fig:1.1} -- possess the property of segmentation when their wave functions can be chosen as
real. This property consists in the appearance of nodes along the periodic coordinate, in such a way
that the distance between them is exactly half a wavelength. Their existence is guaranteed by the
superposition of counter propagating Bloch waves in real combinations, which are supported in turn
by Kramers degeneracy if the system is invariant under time inversion. These nodes disconnect
effectively each segment from the rest of the array and produce Dirichlet boundary conditions on the
finite domains. Although the length of each segment is determined by a prescribed wave number of the
solution -- Bloch quasi-momentum -- the exact location of the nodes can be chosen in various ways,
providing thus disconnected cells of different types. Then, the additional concept of duality comes
into play when the periodic system is made of bi-partite arrays, as illustrated in fig.~\ref{fig:1}.

\begin{figure}[h]
  \vspace{3mm}
    \includegraphics[scale=0.4]{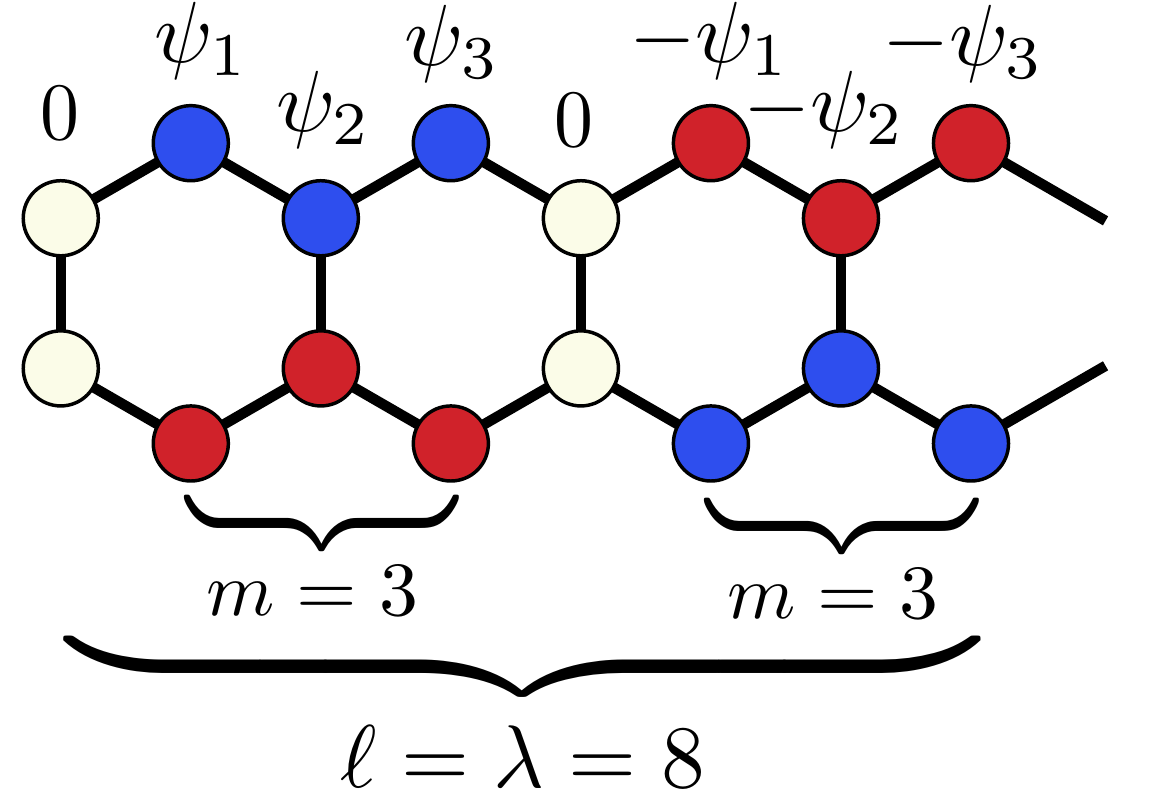}
    \caption{\label{fig:1.2} A real wave $\psi$ on a cycloacene chain. Here, the distance between
      nodes is half the wavelength $\lambda$, which agrees with the system length $\ell$. The length
      of the segments between nodes is denoted by $m$, given in atomic sites.}
\end{figure}

\begin{figure}[h]
  \includegraphics[scale=0.5]{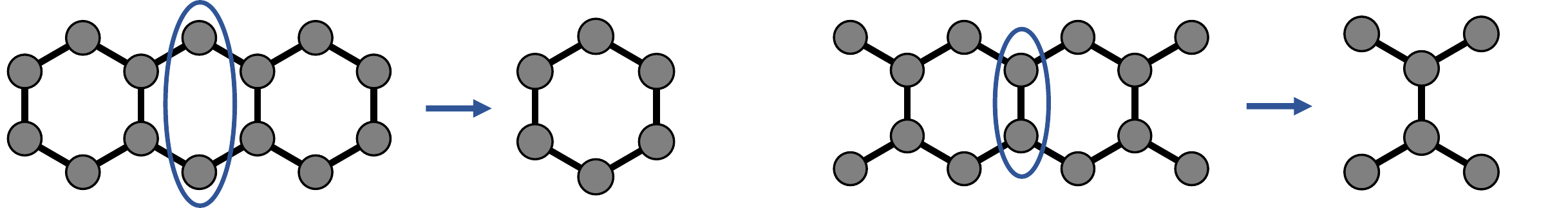}
  \caption{\label{fig:m5} Segmentation of larger blocks into benzene and diallyl.}
\end{figure}

For instance, in a nearest-neighbor tight binding model of cyclopolyacenes, the effective
disconnection is capable of mimicking the electronic $\pi$-orbital of other existing molecules, such
as benzene rings, diallyl, multiple hexagonal rings (polyacenes) or alternative cuts with open rings
such as tetramethyl-naphtalene.%
\footnote{The compound known as durene or durol has the nomenclature 1,2,4,5
  tetramethylbenzene. When the complex involves two central benzene rings, we may refer to it as
  polydurene. The correct nomenclature would be 1,2,9,10 tetramethyl-naphthalene.  See PubChem, URL:
  \url{https://pubchem.ncbi.nlm.nih.gov}} The construction of the wave and the effective
disconnection is shown in fig.~\ref{fig:1.2}.  In fig.~\ref{fig:m5} we illustrate the process of
segmentation into smaller blocks. A moment of thought shows that the wave functions of these blocks
are equally valid solutions of the same global problem, and they correspond to the same energy
provided that segments of different types have energy levels in common. We shall refer to this
property as partial isospectrality among segments. It is important to mention that due to the
periodicity of such solutions, the generated segments are not really independent, so the number of
disconnected regions does not contribute to the degeneracy of the system.

With this in mind, an interesting analogy can be established with the mathematical duality of some
tessellates \cite{sch}. Wall pattern found at the pyramids of Mitla (Mexico) or in greek
architecture can be employed to this end. In fig.~\ref{fig:1} two complementary meanders of type (A)
and (B) are shown. A row can be made equally of AB or BA copies, as one is the complement of the
other. This structure will be examined in the case of organic chains.

\begin{figure}[t]
  \begin{center}
    \includegraphics[scale=0.5]{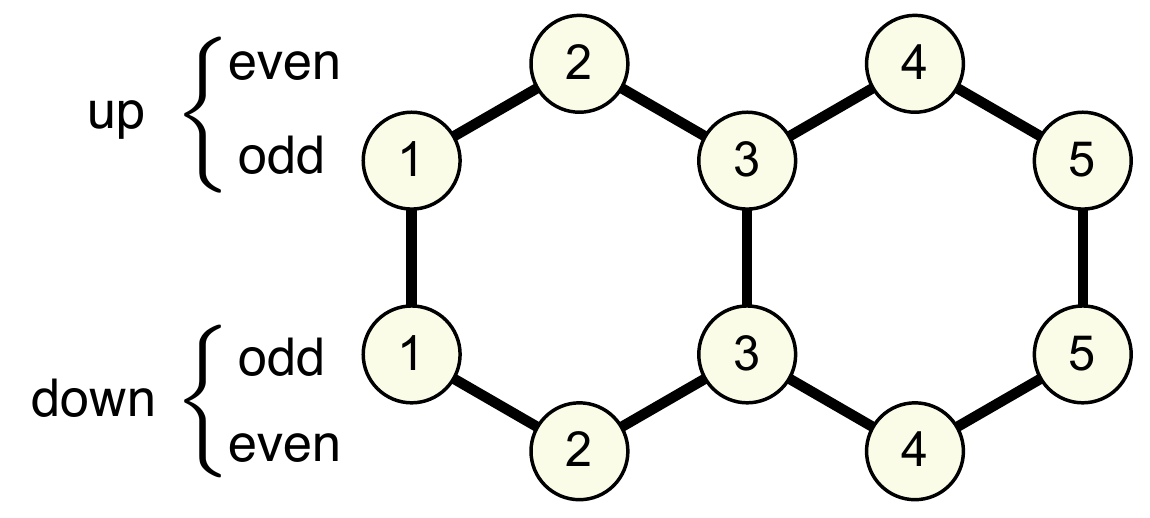}
    \caption{\label{fig:1.3} Decomposition of the array in up and down chains. Inside each chain of
      atoms, one may also decompose the Hilbert space in a direct sum of even and odd numbers. With
      this diagram, we build localized waves ($\pi$ orbitals) denoted by vectors
      $| \mbox{up}, n \>, | \mbox{down}, n \>$ where $n=1,...,2h$.}
  \end{center}
\end{figure}

\section{Duality in organic compounds \label{sec:3}}

One possible way of cutting a hexagonal periodic chain with an even number of hexagons comes from
solutions with a period of 8 carbon atoms. As shown in fig.~\ref{fig:1.2} and~\ref{fig:m5}, this
produces segments of 4 carbon sites (half-wavelength) where the last site of the segment corresponds
to a node, i.e. the segment has an effective length of only 3 sites. We note that the total length
of cyclopolyacene given in carbon atoms is $ \ell = 2h$, where $h$ is the number of
hexagons. Therefore, only when $h = 4 k$ (a multiple of 4) the length $\ell$ is divisible by 8, as
proposed above. Two of these segments have been identified in fig.~\ref{fig:1} as the diallyl
skeleton and the benzene ring. Let us refer to them as AB and BA respectively. For all
nearest-neighbor models of this paper, let us introduce the following scales in the corresponding
Hamiltonian $H$: \bea
\< a| H | b\> &=& 1, \qquad \mbox{if $a,b$ are nearest neighbors,} \nonumber \\
\< a| H | a\> &=& 0, \qquad \mbox{on-site energies are equal.}
\label{1}
\eea
The general form of $H$ for cyclopolyacene is then constructed with the help of a diagram in
fig. \ref{fig:1.3} and it is given by \bea
H= \left\{\sum_{n=1}^{2h}\sum_{\tau = \rm{up, down}} | \tau,(n) \>\< \tau, (n+1) | \right.  
+ \left.  \sum_{n=1}^{h} | \rm{up}, 2n-1 \> \< \rm{down}, 2n-1 \ | \right\} + \mbox{h.c.}
\label{hamiltonian}
\eea
with $(n)\equiv n \mbox{mod} 2h$. The Hamiltonians of the segments AB, BA are built similarly in a
six-dimensional Hilbert space. The BA block, which corresponds to benzene as sketched in
fig.~\ref{fig:1}, has two doublets at energies $\pm1$ and two singlets at energies $\pm 2$, whereas
the AB block, which corresponds to diallyl, has two singlets at energies $\pm 2$, another pair of
singlets at energies $\pm 1$ and finally a doublet at energy $0$.  Since these are equally valid
solutions of the Schr\"odinger equation for the full system, the degeneracy of $E=\pm 1$ is 3, so
there are two triplets in the spectrum of the full array. We should note that the energy at zero is
also doubly degenerate, but this does not follow from Kramers degeneracy, since these states
correspond to a Bloch function with wave vector $\pi$, which coincides with its time reversed
function.

\begin{table}[t]
  \caption{\label{tab:1} Degeneracies of cycloacene. Columns: hexagons. Rows: degeneracies.}
  \begin{ruledtabular}
    \begin{tabular}{lccccccccc}
      Deg$/h$ & 1\footnote{Equivalent to an open chain with 4 sites.} &2&3&4&5&6&7&8&9 \\ \hline
      \\ 1 & 4 &6 &4&4&4 &6 &4&4 & 4
      \\ 2 & 0& 1& 4& 3& 8& 9& 12& 7 & 16
      \\ 3 & 0& 0& 0& 2& 0& 0& 0& 2 & 0
      \\ 4 & 0& 0& 0& 0& 0& 0& 0& 2 & 0
      \\\end{tabular}
  \end{ruledtabular}
\end{table}

\begin{figure}[t]
  \includegraphics[scale=0.7]{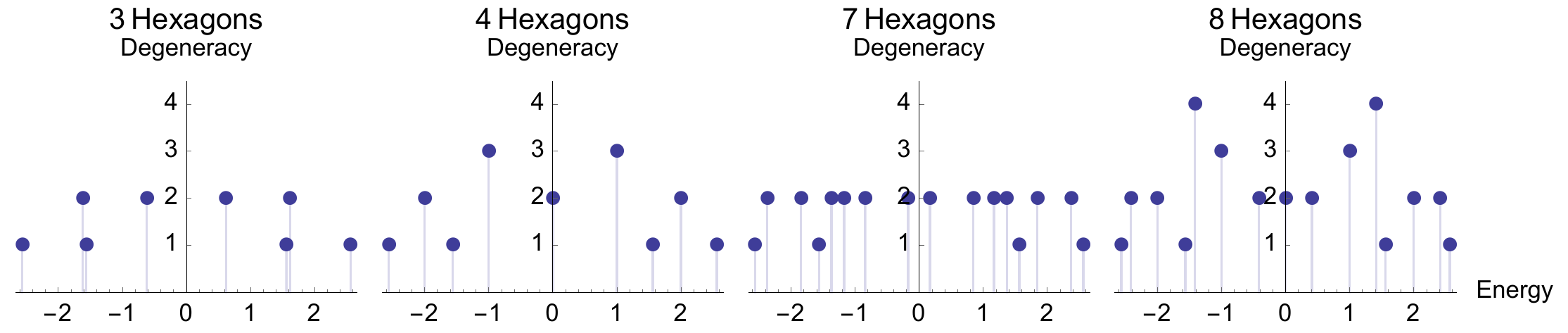}
  \caption{\label{fig:3} Four spectra for the cases of 3, 4, 7 and 8 hexagons in
    cyclopolyacene. While doublets always dominate the spectral set, at $h=4$ we see two triplets at
    $E= \pm 1$ and at $h=8$ two additional quadruplets. The pattern repeats for higher $h$ as
    explained in formulae (\ref{4}) and (\ref{5}).}
\end{figure}

\begin{figure*}[t]
  \includegraphics[scale=0.45]{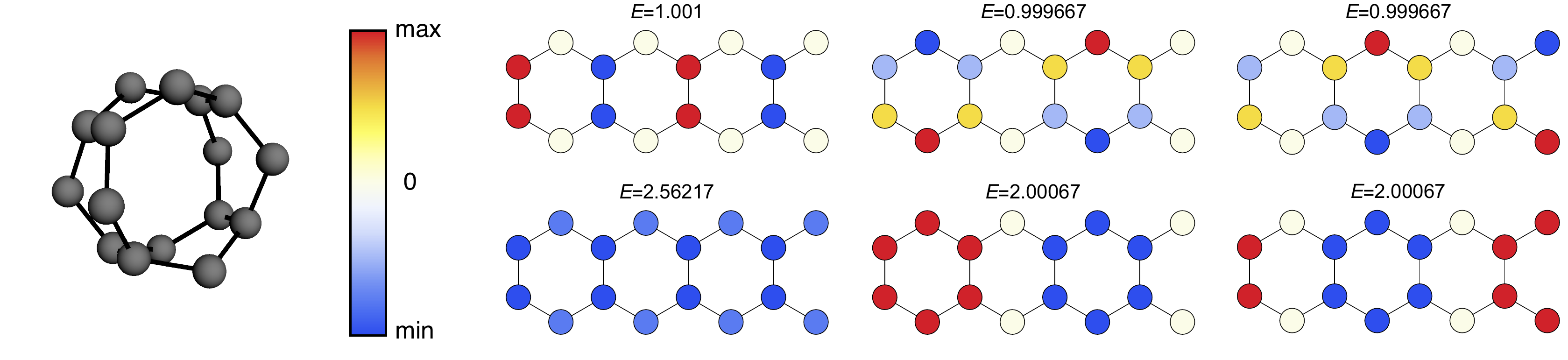}
  \caption{\label{fig:3.1} Numerical wave functions of cyclopolyacene with $h=4$. We show a triplet
    in the first row and a singlet and doublet in the second row. Combinations of triplets may give
    rise to recognizable structures, such as wavefronts propagating to the left or to the
    right. Here we introduce a small perturbation $10^{-3}$ in the vertical couplings, ensuring
    orthogonality of numerical eigenvectors.}
  \vspace{2mm}
  \includegraphics[scale=0.38]{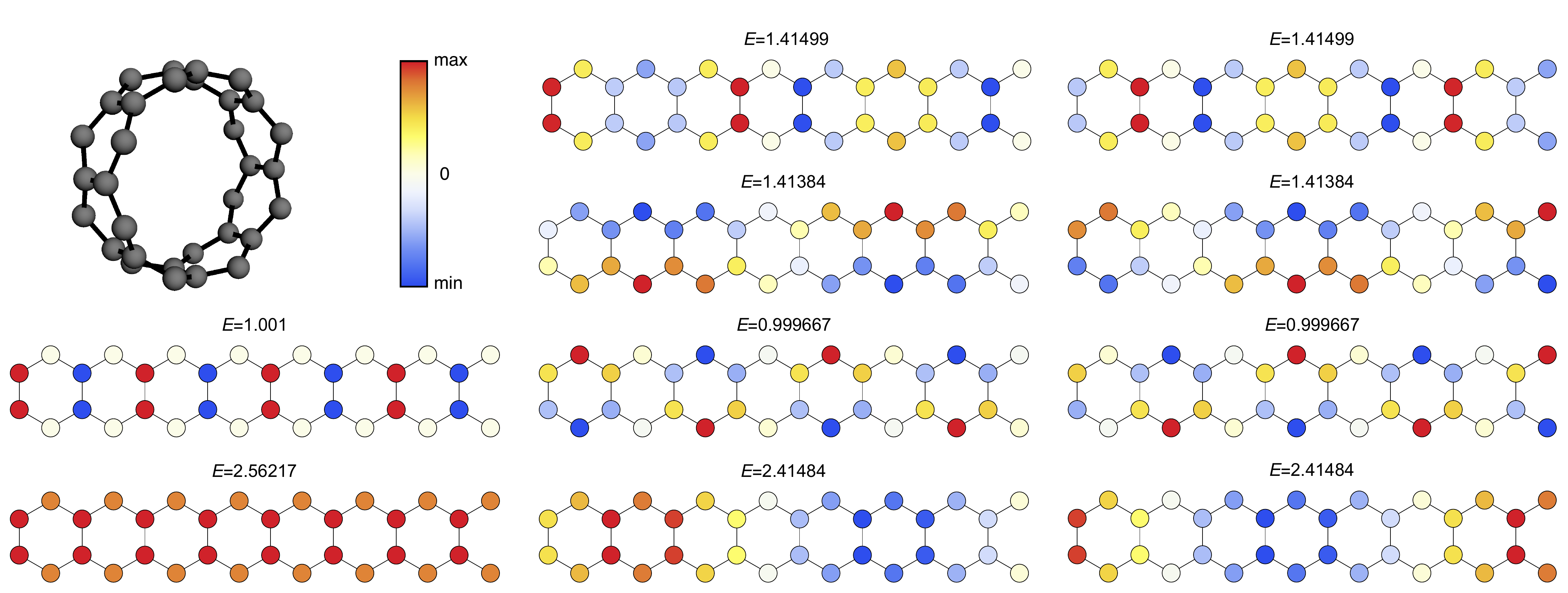}
  \caption{\label{fig:4} Numerical wave functions of cycloacene with $h=8$. We show a quadruplet at
    the top right, a triplet in the third row and a singlet and doublet at the bottom. A small
    perturbation $10^{-3}$ is introduced in the vertical couplings.}
\end{figure*}

When we further increase the length of the chain so as to have $h = 8 k$, we may accomodate
wavelengths of 16 sites and segments of 8 sites, giving an effective segment length of 7 atoms,
corresponding to the molecules known as triacene (3-polyacene or anthracene) and augmented versions
of tetramethyl-naphtalene. These larger segments shown at the bottom of fig.~\ref{fig:1} possess two
doublets at $E = \sqrt{2}$ and $E = -\sqrt{2}$ each, whose union gives rise to 2 quadruplets, each
at $\sqrt{2}$ and $-\sqrt{2}$. As it may appear obvious, the two previously found triplets also
appear here, because each segment of 7 atoms can now be divided into two parts due to the central
node of their antisymmetric wave functions, giving rise again to the AB and BA patterns. A table of
degeneracies is included in tab.~\ref{tab:1} and some cases of interest in fig.~\ref{fig:3}, where
we show the spectra for $h=3,4,7,8$ by way of comparison. In fig.~\ref{fig:3.1} and~\ref{fig:4} we
also show the wave functions for $h=4$ and $8$, respectively. In a comparative manner, we display a
case of triplets, quadruplets, doublets and one singlet of positive energy. It is important to note
that these extraordinarily degenerate waves can be combined to obtain non-trivial nodal
structures. We have thus explained the extraordinary degeneracy of these systems.

In order to characterize completely the multiplets in a molecule of arbitrary length, it is also
important to predict the appearance of singlets as a function of $\ell$. Such singlets are always
limited in number as Kramers degeneracy overrules their existence in the continuous limit of the
cyclic chain, except for the singlet state of vanishing Bloch momentum. Therefore, in the limit
$\ell \rightarrow \infty$ the spectrum must be dominated by all doublets that contain propagating
and counter propagating waves. This implies that the number of singlet states must be bounded
above. In order to count them, we have to show that there must be at least 4 of them regardless of
$\ell$ and, in some cases, we must show an additional pair of singlets emerging only in the sequence
$h=2, 6, 10, ...$ We do this by noting that the 4 compulsory singlets can be generated by applying
transformations to a pre-existing non-degenerate state $|\psi\>$, e.g.  a symmetric state with
vanishing Bloch momentum. This choice of $|\psi \>$ constitutes a nodeless wave of maximal energy
$E_{\rm max}$, in accordance with the positivity of couplings in (\ref{1}). This state can be chosen
to be symmetric with respect to the horizontal line parallel to the chain, which divides the
molecule into upper and lower parts; see fig.~\ref{fig:1.3}. The existence of this symmetric state
is ensured by the the up-down symmetry operator $S$ satisfying $S^2=1$ and $ [S,H]=0$ with the
explicit form %
\bea
S= \sum_{n=1}^{2h} \left\{ |\mbox{up},n\>\<\mbox{down},n|+ |\mbox{down},n\>\<\mbox{up},n|\right\}.
\label{2}
\eea
We have now that $S|\psi\>=|\psi\>$. But from $S^2=1$ there must be another state $|\phi\>$ of zero
Bloch momentum (therefore real) which is antisymmetric $S|\phi\>= - |\phi\>$ and with energy
$E' \neq E_{\rm max}$ due to an increased curvature of the wave with respect to the former
$|\psi\>$. The state $|\phi\>$ is also nodeless on the atomic sites, because there are no atoms
between upper and lower chains. Therefore $|\phi \>$ is also a singlet. In order to find the
remaining two, we note now that the spectrum is symmetric upon inversion with respect to $E=0$. This
specularity property is a consequence of a unitary 'parity' operator $P$ such that
$PHP = -H, P^2 = 1$ and $\{P,S\}=0$, that we now define:
\bea P = \sum_{n=1}^{2h}(-)^n \left\{ |\mbox{up},n\>\<\mbox{up},n|-
  |\mbox{down},n\>\<\mbox{down},n|\right\}.
\label{3}
\eea 
This subtle property pertaining to homogeneous nearest neighbor models (equal sites) is also known
as chiral symmetry in other contexts \cite{beenakker}, and it implies that for any state of energy
$E$ there must be a partner at $-E$. Once more, we choose $|\psi\>$ as the singlet with maximal
energy and clearly $|\bar\psi\>=P|\psi\>$ must be also a singlet; otherwise, using $P^2=1$ would
produce a contradiction. The state $|\bar\psi\>$ produced in this manner is of minimal energy
$-E_{\rm max}$. Now we investigate the symmetry of $|\bar\psi\>$: we have
$S|\bar\psi\> = SP|\psi\>=-PS |\psi\> = -P |\psi\> = -|\bar\psi\>$, i.e. it is antisymmetric. This
means that a symmetric state with reflected energy $-E'$ should also exist, and we can find it by
simply considering $|\bar\phi\>= P |\phi \>$.  Following the same steps, we find that indeed
$S|\bar\phi\> = |\bar\phi\>$. The sought singlets are then
$\{ |\psi \>, |\bar\psi \>, |\phi \>, |\bar\psi \> \}$.

The remaining mystery of the additional pair of singlets occurring at $h=2+4k$ is clarified by
considering waves that accidentally fulfill $P |\psi \> = | \phi \>$ i.e. the phase transformation
$(-)^n$ of the symmetric wave coincides with the antisymmetrized wave. For this to happen, the
on-site wave amplitudes $\<\mbox{up},n |\psi \>$ must be left undisturbed by $P$, while
$\<\mbox{down},n |\psi \>$ must change sign. This is possible only if the components vanish for all
odd $n$: $\<\mbox{up},n |\psi \> = 0=\<\mbox{down},n |\psi \>$. The extraordinary nodal structure of
this state corresponds to a wavelength of two hexagons, i.e. the period is 4 carbon atoms. But now
we argue that this state cannot be degenerate: using Bloch's theorem
$\psi(n) = e^{i \kappa n}\psi_{\rm periodic}$ and the imposed nodal structure, the wave is shown to
satisfy $\psi(n+4)=\psi(n), \psi(n+2)= -\psi(n)$, so we find $\kappa = \pm \pi/2$; also from the
imposed nodal structure, both Bloch momenta $\pm \pi/2$ represent the same wave, because $n$ must be
even for non-vanishing amplitudes and
$\psi(n) = e^{\pm i \pi n/2}\psi_{\rm periodic} = - \psi_{\rm periodic}$ for both values of
$\kappa$. The other singlet emerges by considering the state of opposite parity $S$ again. In
general, a state $|\psi\>$ with all these features is supported by arrays with $h$ even. Once more,
the nodal structure comes to the rescue, as the wave corresponds to disconnected vertical dimers
with energies $\pm 1$ (their Hamiltonian is the Pauli matrix $\sigma_x$) and we have already proved
that for $h=4k$ two triplets arise precisely at these energies, so we must skip all values of $h$
divisible by 4, but not those divisible by 2, and we have finally proved that the sequence
$h=2, 6, 10, ...$ contains 6 singlets.

Our final task is to classify all the states. The number of doublets $o(\underline{2})$ -- here the
symbol $o$ represents the cardinality -- is therefore the remainder after subtracting all triplets,
quadruplets and singlets from the dimension of the system (dim = total number of carbon atoms
$=2\ell = 4h$).  We have the formula \bea o(\underline{2}) &=& (2\ell - o(\underline{1}) - 3
o(\underline{3}) - 4 o(\underline{4}) )/2,
\label{4}
\eea
where
\bea
o(\underline{4}) &=& 2, \quad \mbox{if}\quad h = 8k; \quad 0 \quad \mbox{otherwise}, \nonumber \\
o(\underline{3}) &=& 2, \quad \mbox{if}\quad h = 4k; \quad 0 \quad \mbox{otherwise}, \nonumber \\
o(\underline{1}) &=& 6, \quad \mbox{if}\quad h =2 + 4k; \quad 4 \quad \mbox{otherwise}. 
\label{5}
\eea
Our formula reproduces the numbers in table~\ref{tab:1}. Even a singular case such as $h=1$ is well
captured by (\ref{5}) as $o(\underline{2})=0=(2\times2 -4-3\times 0-4\times 0)/2$. The graph
corresponds to an open chain of 4 sites without degeneracy.

See the Appendix for a detailed exposition of algebraic techniques leading to analytical solutions
for cyclopolyacenes and their constitutive segments.

\section{M\"obius cycloacene \label{sec:4}}

The possibility of synthesizing organic molecules in various configurations \cite{Herges2006} has
led researchers working on conducting polymers to explore a variety of non-trivial geometries.  A
topologically challenging example is the M\"obius strip made of benzene rings. Interestingly, the
electronic spectrum of this non-orientable compound differs considerably from cycloacene. It is
possible to identify important differences in their electronic configurations by analyzing the
degeneracies of the compounds and comparing them as a function of their lengths, measured in atomic
sites. As it is evident, these differences arise solely from the topology of the molecule and,
because of the superposition principle, they produce an important effect on degenerate stationary
electronic configurations. Although we shall not consider specific topological invariants in the
description of such differences (e.g. the Euler characteristic or the homotopy groups) it is
possible to carry out the spectral analysis of all M\"obius cycloacenes and explain their degeneracy
table in terms of symmetry classes of wavefunctions, see table~\ref{tab:m1}.

As in our previous study of orientable cyclic molecules, we shall see here that the method of
segmentation due to periodicity and time-reversal invariance may be applied to the M\"obius strip as
well, but with twice the period as its orientable counterpart.  This allows to identify the
partially isospectral fragments of the chain that occur only for certain lengths. It also provides a
rule for the number of hexagons required to produce triplets in the spectrum.

\begin{figure}[t]
  \includegraphics[scale=0.35]{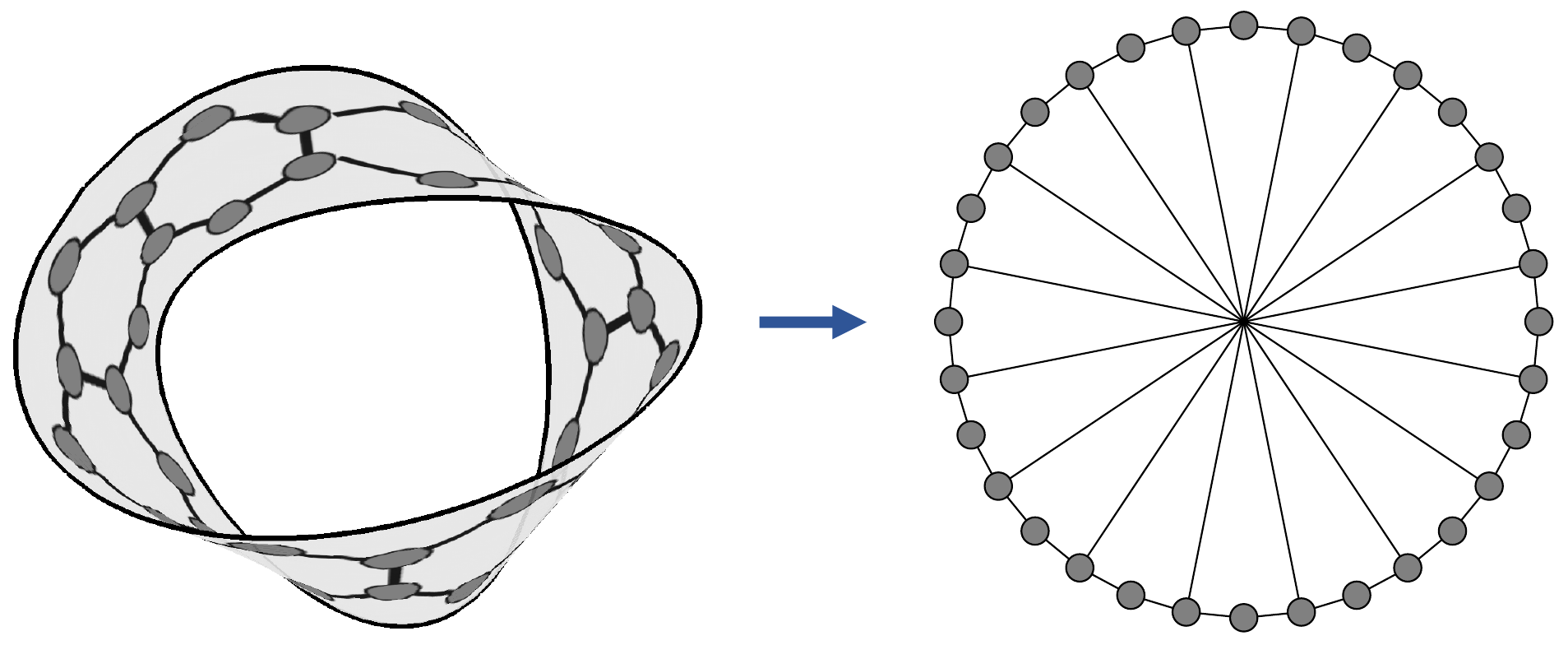}
  \caption{\label{fig:m1} Equivalence between a M\"obius cycloacene and a wheel graph without
    central vertex. All edges represent unit couplings.}
\end{figure}

\subsection{Segmentation of M\"obius cycloacene \label{sec:4.1}}

We start by recognizing that the H\"uckel model of the molecule, which consists of an electron
hopping on a skeletal form, is equivalent to an adjacency graph of interacting sites. The
corresponding nearest-neighbor tight-binding Hamiltonian contains couplings represented by graph
edges, while vertices represent atomic sites of vanishing energies (without loss of generality, if
all sites are identical).

\begin{table}\caption{\label{tab:m1} Degeneracies of the M\"obius cycloacene. Columns:
    hexagons. Rows: degeneracies.}
\begin{ruledtabular}\begin{tabular}{lccccccc}Deg$/h$ & 1 &2&3&4&5&6&7 \\ \hline
\\ 1 & 4 &3 &4&4&4 &3 &4
\\ 2 & 0& 1& 4& 6& 8& 9& 12
\\ 3 & 0& 1& 0& 0& 0& 1& 0
\\\end{tabular}\end{ruledtabular}\end{table}

In fig.~\ref{fig:m1} we show a homeomorphic deformation of a M\"obius strip made of 8 hexagons into
a {\it car wheel\ } graph of 16 sectors, without central site. In the latter, opposite (antipodal)
points on the circle are joined by alternating couplings, representing the common edges of adjacent
hexagons (the pokes of the wheel). This identification comes in handy, as it exhibits the extended
periodicity of the array and justifies the use of cyclic plane waves as a basis of the C$_n$
group. It should be noted though that the full symmetry group of the skeletal form is C$_{2hV}$,
where $h$ is the number of hexagons. The dimension of Hilbert space is $d=4h$.

\begin{figure}[t]
  \begin{center}
    \includegraphics[scale=0.65]{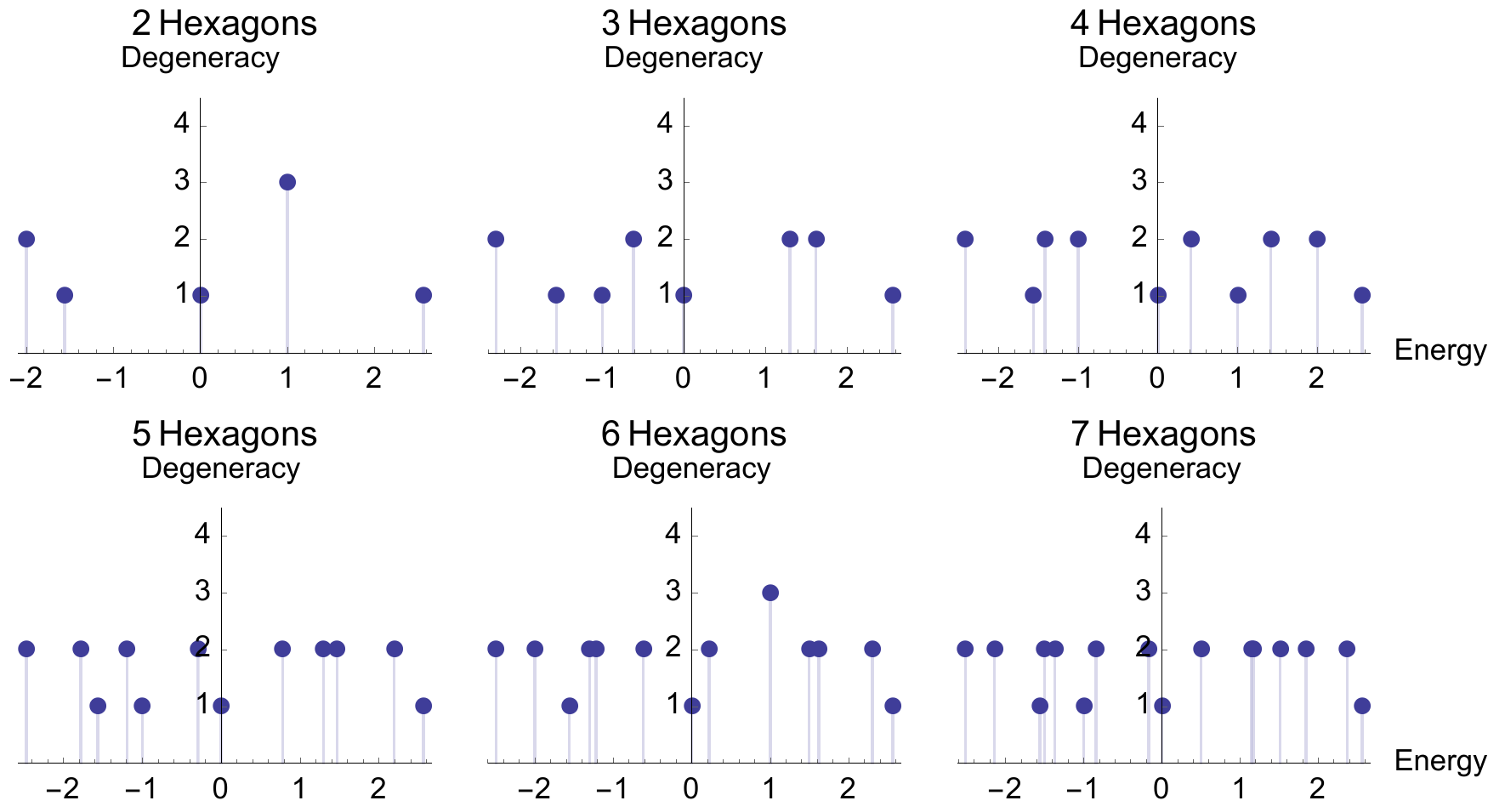}
    \caption{\label{fig:m2} Spectra of various M\"obius cycloacenes. Triplets emerge when the number
      of hexagons $h=4k+2$.  Doublets populate the energy domain as can be seen from Kramers
      degeneracy, and some exceptions arise in the form of singlets.}
  \end{center}
\end{figure}

For definiteness, let us write the Hamiltonian of the M\"obius strip in terms of a localized basis
$|n\>$, numbered by $n=1,...,4h$ around the circle:
\be
H = \left\{\sum_{n=1}^{4h} | (n) \>\< (n{+}1) |
  + \sum_{n=1}^{h}| 2n{-}1 \>\< 2n{+}2h{-}1 | \right\} + \text{h.c.}
\label{m1}
\ee
where $(n)= n$ mod $4h$. The resulting spectra for various values of $h$ are shown in
fig. \ref{fig:m2}. A double degeneracy corresponding to Kramers pairs is visible, while triple
degeneracy emerges only in the cases $h=2, 6, 10, 14 ...$. Equally important is to note that triple
degeneracy occurs at energy $E=1$.

\begin{figure}[t]
  \begin{center}
    \includegraphics[scale=0.25]{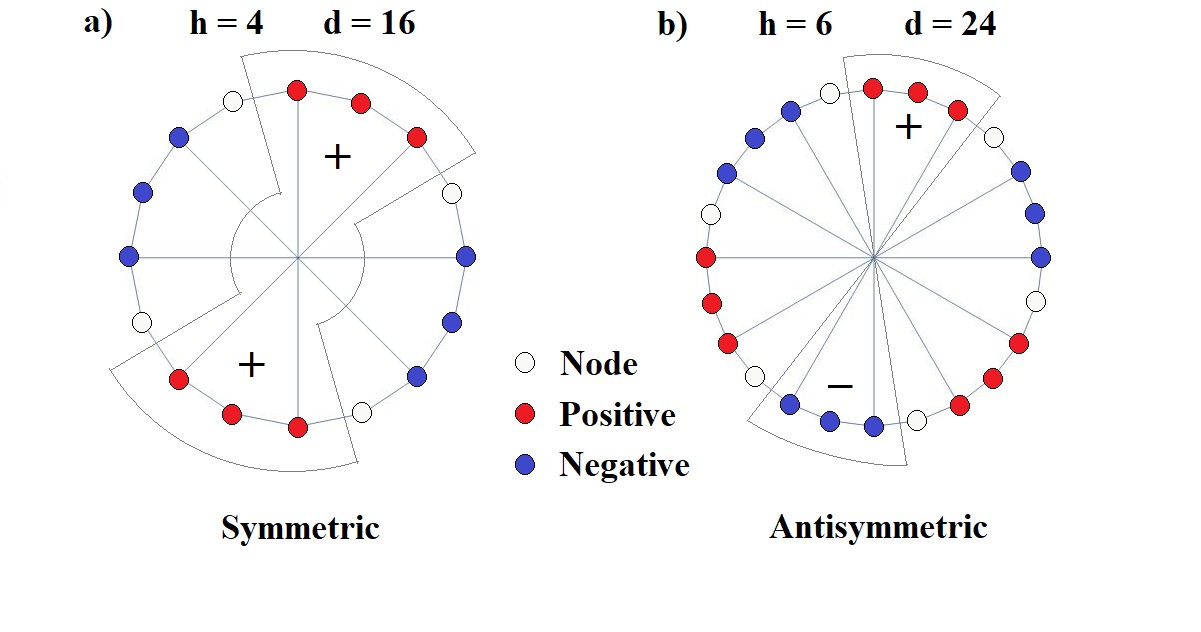}
    \caption{\label{fig:m3} Nodal structure and sign distributions of M\"obius envelope
      wavefunctions (sine and cosine) when mapped onto a circle. In a) we see a case with no
      triplets: the antipodal sectors have the same sign. In b) we see the opposite.}
  \end{center}
  \end{figure}

\begin{figure}[t]
  \begin{center}
  \includegraphics[scale=0.4]{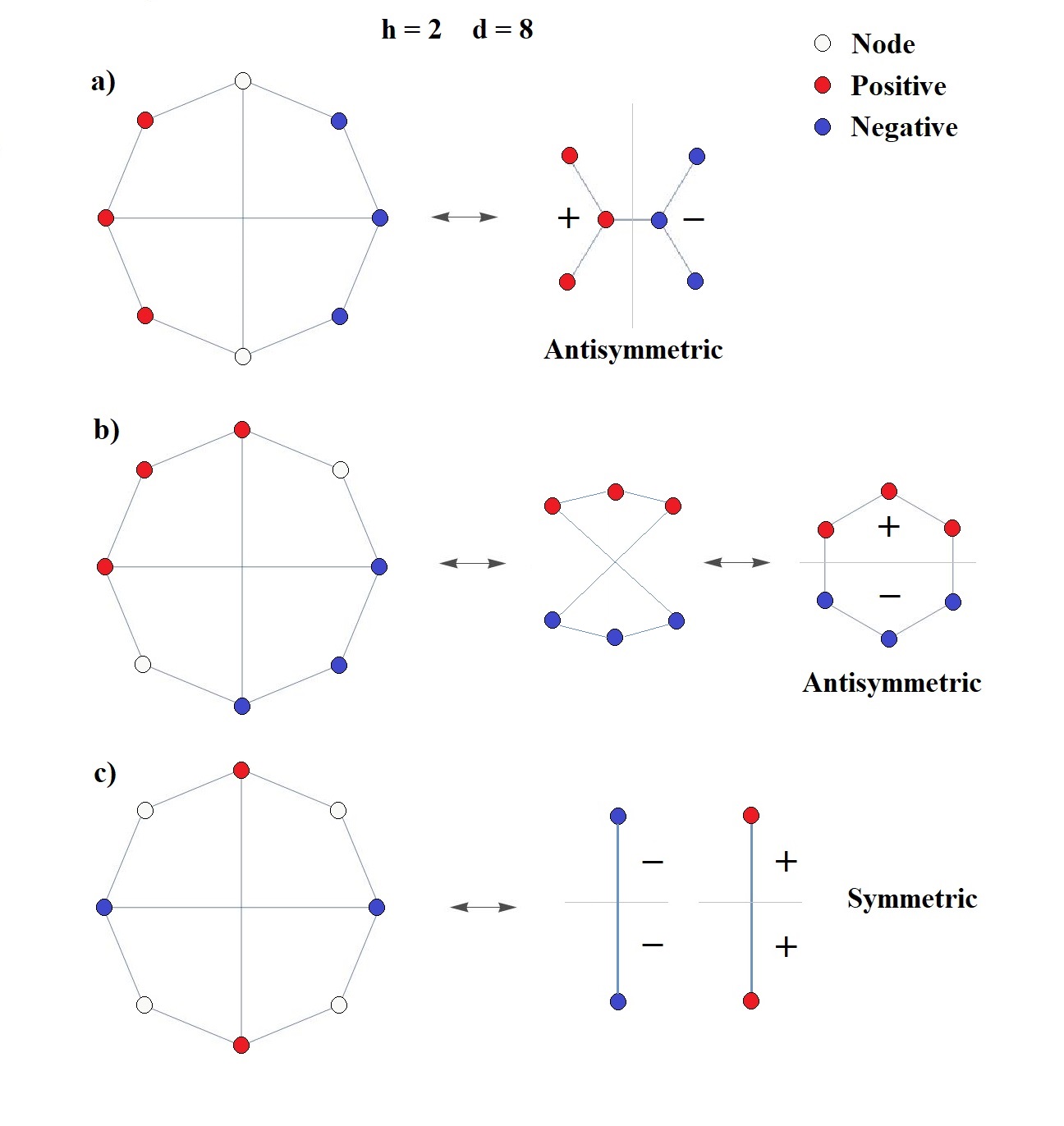}
  \caption{\label{fig:m4} Segmentation of the $h=2$ M\"obius cycloacene, leading to triplets. Panel
    a) diallyl, panel b) benzene and panel c) dimerization.}
  \end{center}
\end{figure}

The possibility of a real cyclic basis offers itself. A solution of the stationary Schr\"odinger
equation can always be written as
\bea
\< n | q \> &=& \exp \left( i \frac{\pi q n}{2h} \right) \< n |\psi \>, \quad q=1,...,2h
\label{m2}
\eea
where the wave $\< n |\psi \>$ has period 2, i.e. $\< n+2 |\psi \>=\< n |\psi \>$. The two possible
choices of $|\psi \>$ together with $2h$ values of $q$ give a set of $4h$ waves. Due to
time-reversal invariance, we may use the set of real and imaginary parts of these waves
%
%
%
\be
\< n | q, c \> = \cos \left( \frac{\pi q n}{2h} \right) \< n |\psi \>, \quad
\< n | q, s \> = \sin \left( \frac{\pi q n}{2h} \right) \< n |\psi \>, \quad q=1,...,2h
\label{m3}
\ee
with the caveat that values of $q \mapsto q+h$ lead to linearly dependent solutions, thus the
dimension is $2 \times 2 \times h =4h$ as expected. Examples of these waves are shown in
fig.~\ref{fig:m3} for $h=4$ and $h=6$, whereby the signs and nodes of the trigonometric envelopes in
(\ref{m3}) are displayed. The nodes occurring at odd or even sites on the circumference correspond to
linearly independent waves. Their wavelength is given by $\lambda=4h/q$ assuming that $q$ divides
$4h$ and the distance between nodes is half a wavelength. A wheel graph has $2h$ sectors; therefore
the largest possible number of nodes is $2h$, corresponding to $q=1$, while for the wave
$\< n | 2h, c \>$ with $q=2h$ the cosine has no zeros. Each of the regions bounded by two nodes is
called a segment.  In the case of a M\"obius strip, we must include in each segment the antipodal
sectors. Because of the coupling between these sectors, each segment corresponds to a graph with two
rows of atoms: upper and lower chains. The length of a segment, i.e. the number of atoms in each
row, is given by $\ell =\lambda/2 -1$ for those values of $q$ that divide $2h$.  It is important to
note that the alternating signs of the segments give rise to different patterns according to the
values of $\lambda$ and $h$: the antipodal sectors will have the same sign if $q=d/\lambda$ is even
and $q$ divides $2h$, corresponding to waves with up-down symmetry. On the other hand, the signs are
opposite if $q$ divides $2h$ but $q$ is odd, leading to up-down antisymmetric functions. An example
of segmentation is shown for the case $h=2$ in fig.~\ref{fig:m4}. In this example, we only show
cases relevant for solutions with $E=1$. Other cases can be done as an exercise.

When the segments are large due to a large wavelength, it is always possible to subdivide them into
smaller ones using symmetric or antisymmetric functions with respect to a vertical symmetry
axis. This was shown in the previous case of cycloacenes in fig.~\ref{fig:m5}. Examples of small
segments are given in fig.~\ref{fig:m4}, where a length of 3 atoms produces the skeletal forms of
diallyl in panel a) or benzene in panel b). The minimal segment of 1 atom leads to full
dimerization of the array, as shown in fig. \ref{fig:m4}, panel c). The subdivision process always
corresponds to a new choice of smaller wavelength with the same nodal position.

For completeness, let us write the Hamiltonian of a polyacene segment. As in cycloacenes, there are
two types: closed hexagons at the ends of the chain and 2 open leads at both ends of the chain (in
chemistry, this would be a tetramethyl group attached to the ends of anthracene). The convenient
re-ordering of site numbers shown in fig.~\ref{fig:1.3}, upper and lower panels, can also be employed
here. For the closed hexagon case we have
\be
H_1 = \left(\begin{array}{cc} K &  \Delta \\  \Delta & K  \end{array} \right) \quad \text{with} \quad
\Delta = \mbox{diag} \left\{ 1,0,1,0,...1,0,1 \right\}, \quad
K = \sum_{n=1}^{\ell} |n\>\<n+1| + \mbox{h.c.}
\label{m4}
\ee
and for the chain with open leads
\be
H_2 = \left(\begin{array}{cc} K &  \tilde \Delta \\  \tilde \Delta & K  \end{array} \right), \quad
\text{with} \quad \tilde \Delta = \mbox{diag} \left\{ 0,1,0,1,...0,1,0 \right\}.
\label{m5}
\ee

\begin{figure}[h]
  \includegraphics[scale=0.5]{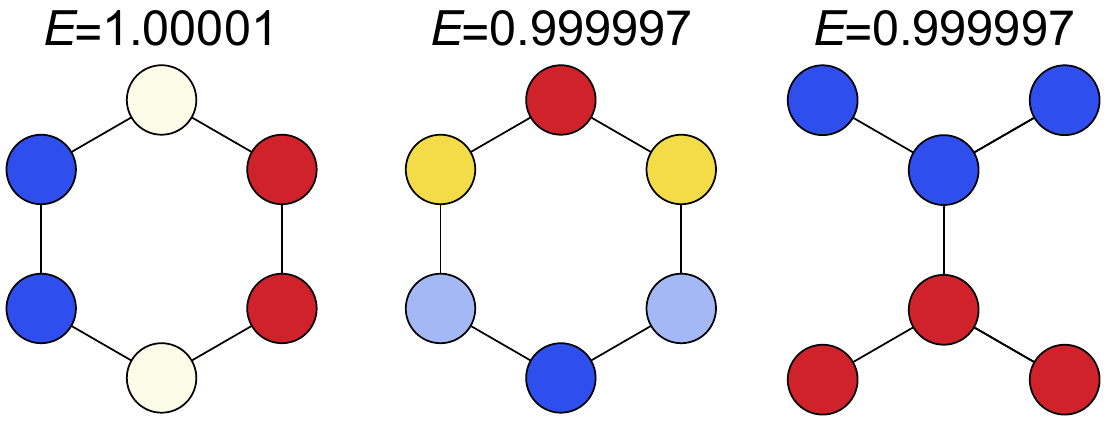}
  \caption{\label{fig:m7} Numerical wave functions for $E=1$, corresponding to accidental
    triplets. Artifacts may arise when exact degeneracy is present, so we introduce here a small
    perturbation $10^{-5}$ in the vertical couplings in order to obtain numerically orthogonal
    functions. Actual energies differ by the same amount, as indicated at the top of each plot. Left
    panel: left-right antisymmetry and dimerization of benzene. Central panel: up-down antisymmetry
    for benzene. Right panel: up-down antisymmetry for diallyl.}
\end{figure}

\begin{figure}[h]
  \includegraphics[scale=0.45]{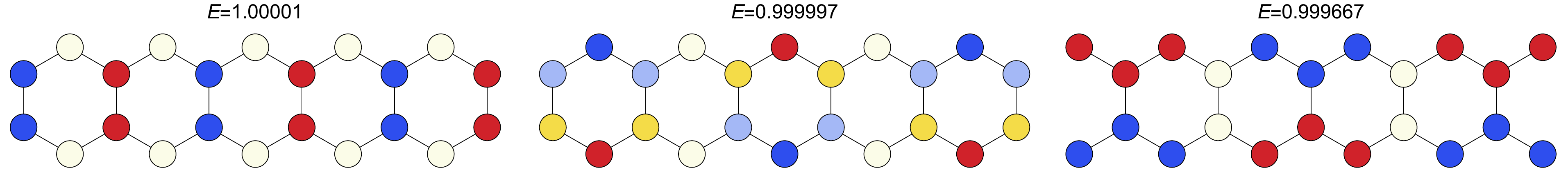}
  \caption{\label{fig:m8} Wave functions for large segments, corresponding to degenerate states
    $E=1$. As in fig. \ref{fig:m7}, we find dimerization patterns (top panel), benzene
    antisymmetric functions (central panel) and diallyl antisymmetric functions (bottom panel).}
\end{figure}

The stationary wave functions of these Hamiltonians correspond to those of the segments conforming
the M\"obius band, as long as they belong to the allowed up-down symmetry class discussed
above. Some plots are shown in fig. \ref{fig:m7} (segments of length 3 and six sites, diallyl and
benzene) and fig. \ref{fig:m8} (length 11 and 22 sites, anthracenes).  These waves correspond to the
degenerate energy triplets $E=1$ that we discuss in the following.


\subsection{M\"obius degeneracy: triplets \label{sec:4.2}}

Now we focus on the appearance of triplets for $h=4k+2$. In order to rule out the case of $h$ odd,
we note that the segmentation process does not lead to minimal blocks such as those in
fig.~\ref{fig:m5}, therefore the solutions of this problem may contain the energy $E=1$ only when
there is full dimerization of the compound, i.e. when the nodes of the wave function display a full
disconnection of vertical dimers. However, the antipodal sign distribution imposes an antisymmetric
function for such dimers. Since each dimer is described by a Hamiltonian given by $\sigma_x$, the
energy $E=1$ is ruled out and $E=-1$ is justified, but only as a singlet.

The case $h=4k$ can be discarded as well, using the antipodal sign structure of the wave functions
as before: the antipodal sectors of twisted benzene rings (shown in panel (a) of fig.~\ref{fig:m3})
necessarily possess equal signs in this case.  However, one of the $E=1$ solutions of an isolated
benzene ring is antisymmetric with respect to the horizontal axis, as shown in fig.~\ref{fig:m7},
central panel. This forbids the existence of the three solutions shown in fig.~\ref{fig:m7}
simultaneously in the segments of this array; since there are no other waves with $E=1$, triplets
are ruled out. Also, the state at $E=1$ must be a singlet because, although the full dimerization
allows such solution, the diallyl block in the right panel of fig~\ref{fig:m7} is not allowed to
have up-down antisymmetry. This can be easily checked by translating one site the nodes in panel (a)
of fig.~\ref{fig:m3}. Therefore, only the symmetric diallyl solution is allowed, with an energy
$E \neq 1$. This elimination leaves us with the dimer solution as a singlet.

Finally, the degeneracy table \ref{tab:m1} is explained by recognizing that either $E=1$ is a
triplet or a singlet, as the dimer state coalesces with the benzene and diallyl doublet in
fig. \ref{fig:m7} when $h=4k+2$. This means that the number of singlets is diminished by 1 for this
sequence of $h$, leading to $4-1 = 3$ singlets. In the other cases, it is easy to see that 4
singlets must appear as a consequence of the group C$_{2h}$, whose eigenphases $1$ and $-1$ do not
have complex-conjugate companions. Using the cyclic waves (\ref{m2}), we see that there must be two
solutions for each eigenphase (Bloch momentum) built from the basis of even and odd sites
$\< n | q\>, \< n+1 | q\>$. Hence 4 singlets.

\section{Cycloacene degeneracies derived from the continuous limit}

The rather intricate dependence of the accidental degeneracies discussed above on the molecular
length, while it can be grasped by a careful and detailed study of the wave functions as described
above, nevertheless does not leave one with a clear overall picture of what is happening. In the
following, we show that such a picture can be obtained for the fully infinite system: if we study
the infinite polyacene chain, with no boundary conditions, we are led to a continuous spectrum. This
spectrum depends on a continuous parameter, the quasimomentum $k$ as well as a discrete parameter
$\tau=\pm1$, which describes the parity symmetry with respect to the horizontal axis.  In this case,
one shows that each eigenvalue belonging to $k\neq0,\pi$ outside of a given interval, is {\em
  fourfold degenerate}: first, trivially, there is a degeneracy between $k$ and $-k$, for the same
value of $\tau$. Second, however, for all values of $k$ outside the interval
$\pi/2\leq|k|\leq2.16706$, to every $k_1$ and $\tau$ there exists a corresponding value $k_2$ such
that the energy corresponding to $k_1$ and $\tau$ is equal to that corresponding to $k_2$ and
$-\tau$.

This result obtains as the result of a simple computation detailed in the following subsection. If
we now ask how this translates to the finite case, we observe the following: for a degeneracy to
arise in any of the finite cycloacene chains, say of length of $h$ hexagons, it is clearly necessary
that both $k_1$ and $k_2$ lead to eigenfunctions having periodicity $h$, so that both $k_1$ and
$k_2$ must be rational multiples of $2\pi$. As it turns out, this condition cannot be satisfied
except in very specific cases for values of $k=\pi, \pi/2$ and $\pi/4$. Therefore the extreme
specificity of the degeneracy scheme for cycloacenes arises from the combination of two effects: (i)
the existence of a very general degeneracy in the continuous case, valid for a continuum of
eigenfunctions, and (ii) the fact that none of these pairs of degenerate eigenfunctions in the
continuum are both compatible with the periodic boundary conditions for any finite chain, except for
a few quite exceptional cases.

\subsection{Infinite cycloacenes and their degeneracies}

Let us hence consider first an infinite chain of hexagons, that is, we do not impose {\em any\/}
boundary conditions, but rather use explicitly the full infinite system. In this case, the spectrum
is continuous and the eigenfunctions are Bloch functions with quasimomentum $k$. These have the
property that their value on any site is equal to $e^{ik}$ times the value they take on the
corresponding site of the hexagon to the left. One readily shows, see the Appendix, that the
Hamiltonian \eqref{hamiltonian} limited to such functions can be reduced to
\begin{eqnarray}
  H(\tau, k)=\frac\tau2\left(
                {\mathbb1}-\sigma_z
                \right)+(1+\cos k)\sigma_x+\sin k\sigma_y 
            =\frac\tau2{\mathbb1}+\vec{e}_{\tau,k}\cdot\vec\sigma
\end{eqnarray}
with
\begin{equation}
  \vec{e}_{\tau,k}=\left( 1+\cos k,\sin k,\frac\tau2 \right).
\end{equation}
Here the $\sigma$'s are the Pauli matrices, and $\vec\sigma$ the vector of Pauli matrices, and
$\tau=\pm1$ corresponds to the parity of the function under the operation $S$ described in
\eqref{2}.

Clearly the eigenvalues of $H(\tau,k)$ correspond to $\tau/2\pm||\vec{e}_{\tau,k}||$. The issue of
degeneracy thus corresponds to finding pairs of values $(\tau_1,k_1)$ and $(\tau_2,k_2)$ such that
the values of these norms lead to degeneracies. For $\tau_1=\tau_2$ nothing of interest arises,
whereas for $\tau_1\neq\tau_2$ we are led to the condition
\begin{equation}
||\vec{e}_{\tau_1,k_1}||-||\vec{e}_{\tau_2,k_2}||=1.
\label{eq:degen}
\end{equation}
To reduce this to polynomial form, we introduce
\begin{equation}
x_i=\cot^2\left(\frac{k_i}2\right)\qquad(i=1,2).
\label{eq:defx}
\end{equation}
After some tedious algebra, (\ref{eq:degen}) yields
\begin{equation}
  2 ( x_1-x_2)^2   
  -2 x_1^2 x_2^2-3( x_1
  x_2^2+ 
  x_1^2 x_2) -(x_1^2+x_2^2)-4 x_1
  x_2-(x_1+x_2)=0
\label{eq:poly}
\end{equation}
Thus, for every $x_1$ there are, in principle, two values of $x_2$ leading to degeneracy. Since we
need {\em positive\/} real roots, see (\ref{eq:defx}), we see after a bit of algebra that
(\ref{eq:poly}) has exactly one positive real root $x_2$ if $x_1\in{\cal C}$, where
\begin{equation}
  {\cal C}=\left\{x:0<x<\frac14\left(\sqrt{17}-3\right)\mbox{\rm{} or }x>1\right\}
\end{equation}
with $(\sqrt{17}-3)/4\approx0.2808$, and none otherwise. Since the equation is symmetric in $x_1$
and $x_2$, we see that for $x_1\in{\cal C}$, we have one value of $x_2\in{\cal C}$ satisfying
(\ref{eq:poly}).  Finally, looking at (\ref{eq:defx}), we see that the value of $x_{1,2}$ only
defines $k_{1,2}$ up to sign.  This is natural, since $\pm k$ always correspond to degenerate
energies, except when $k=0$ or $k=\pi$.  It follows therefore that for the corresponding values of
$k$ the continuous spectrum of the infinite polyacene chain is fourfold degenerate for all values of
$k$, with the 4 values of the form $\pm k_1$ and $\pm k_2$.  This corresponds to the whole range of
$k$ except $\pi/2\leq|k|\leq2.16706$.

Now, in order for this to yield any kind of degeneracy on a finite cycloacene, we must satisfy the
additional conditions
\begin{equation}
  \exp\left( ihk_1 \right)=1, \quad
  \exp\left( ihk_2 \right)=1,
  \label{eq:period}
\end{equation}

where $h$ is the length (measured in hexagons) of the polyacene chain. In other words, both $k_1$
and $k_2$ must be rational multiples of $2\pi$. This is a strong condition, and the only solutions
we are aware of are the following
\begin{subequations}
  \begin{eqnarray}
    &k_1= \pi,\qquad &k_2=\pi,\\
    &k_1= \pi,\qquad &k_2=\pi/2,\\
    &k_1= \pi/4,\qquad &k_2=3\pi/4
  \end{eqnarray}
  \label{eq:solspec}
\end{subequations}
and those which arise by change of sign. Note that $k_1=\pi$ is exceptional, as it corresponds to 2
different values of $k_2$: this corresponds to the singular case $x_1=0$, in which double zeros
arise.

The conditions (\ref{eq:period}) can be transformed to polynomial conditions in the $x_{1,2}$,
yielding
\begin{equation}
  T_h\left( \frac{x_1-1}{1+x_1} \right)=1, \quad
  T_h\left( \frac{x_2-1}{1+x_2} \right)=1,
\label{eq:period1}
\end{equation}

where $T_M(x)$ is the $h$'th Chebyshev polynomial. It is an interesting question whether any
solutions to (\ref{eq:poly}) satisfying (\ref{eq:period1}) could be found, apart from those listed
in (\ref{eq:solspec}).

As an amusing possible extension, note that if we vary the coupling constants assuming that the
interactions between up and down atoms have size $1$ and those between atoms of the same layer have
size $v\in\mathbb{R}$, (\ref{eq:poly}) becomes
\begin{equation}
2 v^2( x_1-x_2)^2   
-2 x_1^2 x_2^2-3( x_1
x_2^2+ x_1^2 x_2)- (x_1^2+x_2^2)-4 x_1
   x_2-(x_1+x_2)=0
\label{eq:polyv}
\end{equation}
It is then straightforward, starting from values of $k_{1,2}$ satisfying (\ref{eq:period}),
algebraically to determine values of $v$ for which (\ref{eq:polyv}) is satisfied. This provides
special values of $v$ for which cycloacenes of length $h$ have an (accidental) degeneracy. Thus for
$h=5$, that is, polyacenes with length an integer multiple of $5$, a fourfold degeneracy arises for
$v=\sqrt{6/5}$, for the eigenvalues $\pm3/\sqrt5$. This specific result is found by taking
$k_1=2\pi/5$, $k_2=6\pi/5$, calculating $x_1$ and $x_2$ using (\ref{eq:defx}) and evaluating $v$ via
(\ref{eq:polyv}).

\section{Conclusions \label{sec:5}}

We have explained the existence of degeneracy in periodic systems using the following key
ingredients: i) effective segmentation supported by real periodic waves and time reversal invariance
ii) hidden duality in the construction of tilings and, less generically but equally valid iii)
partial isospectrality of building blocks. In the chemical application presented here, these
concepts translate into i) Bloch waves (or their polygonal reductions) surrounding cyclopolyacenes
ii) a hidden duality featured by benzene and diallyl, as well as polyacene and
tetramethyl-naphtalene, and iii) their partial isospectrality. With these, we were able to predict
the nature of all degeneracies in the electronic structure of arbitrarily long hexagonal chains,
specialized of course to $\pi$-orbitals.

An important mathematical conclusion springs immediately from our discussion: While it is well
established that accidental degeneracies can be reduced to internal unitary symmetry groups acting
on Hilbert space, the true meaning of hidden symmetries lies in the possibility of transforming
physical systems {\it beyond\ } configuration space, e.g. phase space. So, is cyclopolyacene a case
of hidden symmetry? The answer must be positive, as the permutational group acting on the space of
carbon atoms does not explain the phenomenon by itself. Here, the duality provided by segmentation
rests on the choices of Bloch's momenta of certain wavelengths. The role played by the momentum
(i.e. the generator of translations along the molecule, which are equivalent to molecular rotations
around the axis perpendicular to its centre) and the hexagonal symmetry of the building blocks point
to a property that may be associated to a discrete phase space, rather than the mere configuration
of the chemical compound.

The implications on the construction of other systems are now obvious. Although it is not easy to
engineer new molecules based on these ideas \cite{sadurni2020}, the spectral structure can always be
replicated by design in quantum wires \cite{appenzeller, sadurni2016}, quantum graphs
\cite{qg1,qg2,qg3}, microwave \cite{proximity, dmo, s2010, stegmann2017, stegmann2017a, stegmann2020} and
acoustic \cite{ramir2019} systems based on the tight-binding paradigm.

\begin{acknowledgments}
  The authors gratefully acknowledge financial support from CONACYT Proyecto Fronteras 952, Proyecto
  A1-S-13469, and the UNAM-PAPIIT research grants IA103020, IN113620 and AG100819. We thank
  Hans-J\"urgen St\"ockmann for his invaluable comments.
\end{acknowledgments}

\section*{Data availability}
The data that supports the findings of this study are available within the article.

\appendix

\section{Dirac description of cyclopolyacene}

In full analogy with the application of Dirac operators in graphene, boron nitride \cite{s2010} and
linear chains - e.g. the emulation of a Dirac oscillator \cite{dmo} -- we introduce here a formalism
of $4\times 4$ matrices for long molecules made of carbon rings. Exact energies and eigenfunctions
\cite{skarmakar} shall be obtained with little effort.  We start with appropriate definitions of
operators. For localized states in the $\pi$ orbital of constitutive carbon atoms we use the kets
$|\mbox{up}, (n) \>, |\mbox{down},(n)\>$ with $n=1,...,2h$ and $(n)=n \mbox{mod} 2h$ and $h$ the
number of hexagons. Let the Hamiltonian be
\begin{equation}
  H = \sum_{n=1}^{2h} \left[ | \mbox{up}, (n) \>\< \mbox{up}, (n{+}1)  |
        + | \mbox{down}, (n) \>\< \mbox{down}, (n{+}1)  | \right]
  +  \sum_{n\, \mbox{\small odd}} | \rm{up}, (n) \>\< \mbox{down}, (n)  | + \rm{h.c.}
\label{s1}  
\end{equation}
where h.c is the hermitian conjugate of the full expression at the left. A more tractable matrix can
be obtained if we employ the re-ordered basis $ | i,j,m \> = | i \> \otimes | j \> \otimes | m \> $,
where $m=1,...,h$ runs over the hexagonal rings; $i=1,2$ denotes up and down chains respectively,
and $j=1,2$ indicates odd and even sites in each chain. For example, the first site of the upper
chain has the state $|1,1,1\>$, the second site of the upper chain $|1,2,1\>$, the third site of the
upper chain $|1,1,2\>$, and so on. This allows to employ the following Pauli matrices:
\bea
\sigma_{+} &=& \v 1_{2\times 2} \otimes |1\>\<2|, \quad \sigma_{-}
= \sigma_{+}^{\dagger}, \quad [ \sigma_{+} , \sigma_{-} ]= 2\sigma_3 \nonumber  \\[2mm]
\tau_{+} &=& |1\>\<2| \otimes \v 1_{2\times 2}, \quad \tau_{-}
= \tau_{+}^{\dagger}, \quad [ \tau_{+} , \tau_{-} ]= 2\tau_3 \nonumber  \\[2mm]
\sigma_1 &\pm&  i\sigma_2 = \sigma_{\pm}, \quad \tau_1 \pm i\tau_2 = \tau_{\pm}, \quad [ \tau_i , \sigma_j ] = 0.
\label{s2}
\eea
We also need a kinetic operator that contains translations along hexagons, given by
\bea
K = \sum_{ m=1}^{h} | [m]\>\<[m] | + | [m]\>\<[m+1] |,
\label{s3}
\eea 
where $[m] = m \, \text{mod} \, h$, which provides periodic boundary conditions by coupling the last and the
first hexagonal cells. These definitions pay off, as the Hamiltonian (\ref{s1}) acquires the Dirac
structure
\bea
H = \sigma_{+} K + \sigma_{-} K^{\dagger} + \frac{1}{2}\left( \v 1_{4\times4} + \sigma_3 \right) \tau_3.
\label{s4}
\eea
As in the Dirac equation, the last term can be interpreted as a mass term or spectral gap, while the
first two terms are the kinetic energy with momenta
$K_1 = K + K^{\dagger}, K_2 = i (K - K^{\dagger})$ and Dirac matrices $\alpha_i = \sigma_i$,
$\beta = \sigma_3 \tau_3$. For the sake of clarity, the explicit matrix for $H$ is included here:
\be
H = \left(
  \begin{array} {cccc}
    0 & K & \v 1 & 0 \\
    K^{\dagger} & 0 & 0 & 0 \\
    \v 1 & 0 & 0 & K \\
    0 & 0 & K^{\dagger} & 0 \end{array}  \right),
\label{s5}
\ee
where the operator $\v 1$ is the $h \times h$ identity matrix. The spectrum and eigenfunctions are
easily obtained by finding the common eigenvectors of $K, K^{\dagger}$, as these operators
commute. Using the system's periodicity, the finite dimensional analogues of Bloch waves are given
by
\bea
| k ) \equiv \frac{1}{\sqrt{h}} \sum_{m=1}^{h} e^{ikm} | m \>, \quad K | k) = (1 + &&e^{ik}) | k ),
\label{s6}
\eea
with $k=2\pi q/h$ and $q=1,\hdots,h$. These waves can be used to our favor, since $H$ can be reduced
now to $4\times4$ blocks $H^{(k)} = (k | H |k )$ given below
\be
H^{(k)} = \left(
  \begin{array} {cccc}
    0 & 1 + e^{ik} &  1 & 0 \\
    1 + e^{-ik} & 0 & 0 & 0 \\
    1 & 0 & 0 & 1 + e^{ik} \\
    0 & 0 & 1 + e^{-ik} & 0
  \end{array}  \right).
\label{s7}
\ee
The form of the spectrum and eigenfunctions contains now
$k = 2 \pi / h, 2(2 \pi / h),3(2 \pi / h), ..., 2\pi$, $s=\pm 1$ and $t=\pm 1$ as quantum numbers;
we have
\bea
E_{k,s,t} = \frac{t + s \sqrt{ 1 + 16 \cos^2 (k/2) }}{2}, 
\label{s8}
\eea
and with the aid of the rotated vectors ($t= \pm 1$, $j=1,2$)
\bea
| t \> \equiv \frac{1}{\sqrt{2}} \left( |1\> + t | 2 \> \right)&,& \quad
| j, \pm \> \equiv \hat R( \phi, \theta_{\pm}) | j \>, \nonumber \\[2mm]
\hat R(\alpha,\beta) = \exp (-i \tau_3 \alpha/2) \exp (-i\tau_2 \beta/2), \quad
&&\tan(\theta_{\pm}) = \pm 4 |\cos(k/2)|, \quad \tan(\phi)= - \frac{\sin k}{1 + \cos k},
\label{s9}
\eea
the eigenfunctions read
\bea
| k, s, t\> = | t \> \otimes | s, t \> \otimes | k ).
\label{s10}
\eea
The explicit form of the rotated spinors $| s, t \>$ is fairly easy to compute in terms of the
angles $\phi$, $\theta$ \cite{skarmakar} with the help of the exponentials conforming $\hat R$
above.

\begin{figure*}[t]
\begin{center}
\includegraphics[scale=0.53]{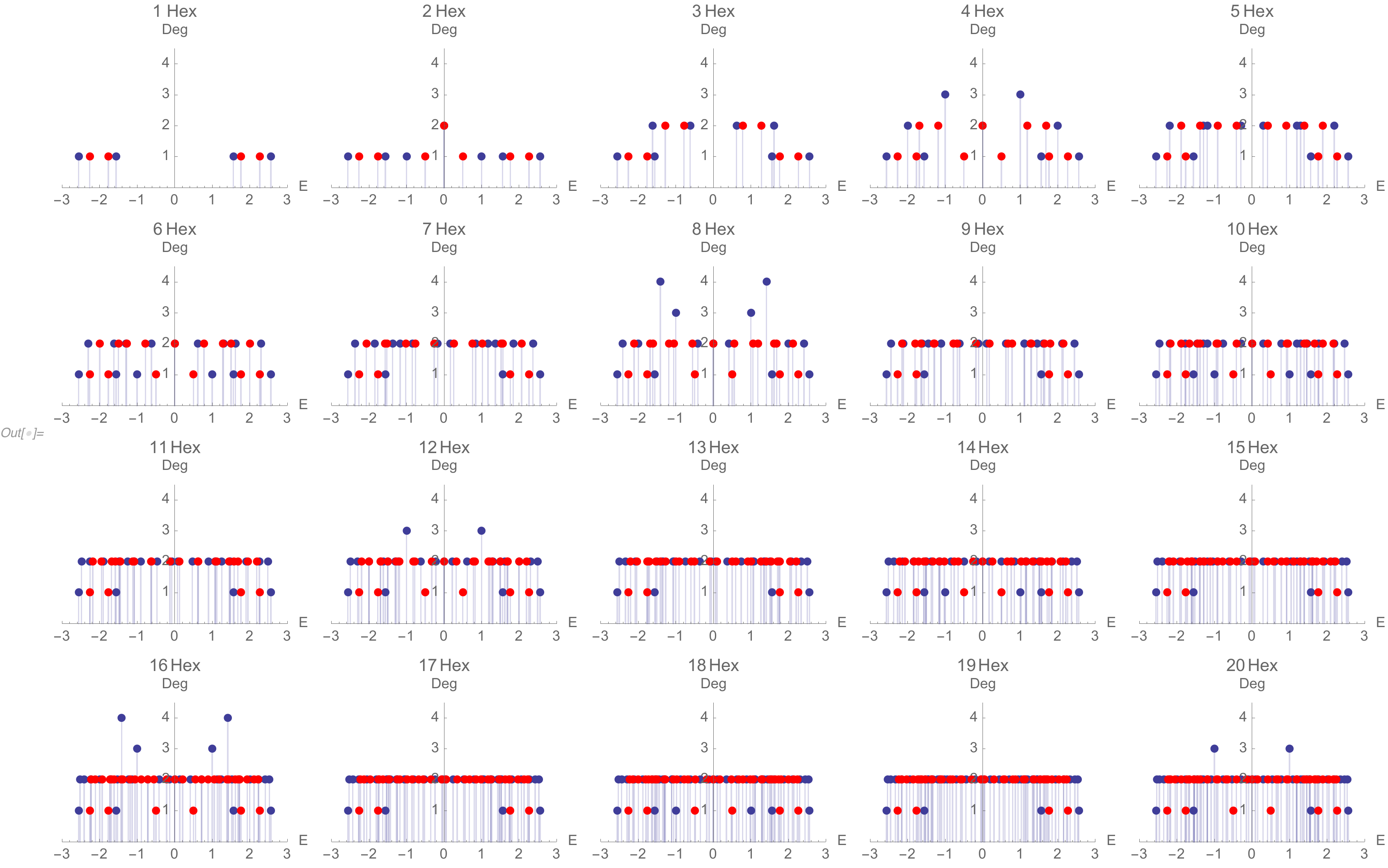}
\end{center}
\caption{\label{fig:s1} Energies without symmetry breaking (blue dots) and with $\Delta = 0.5$ (red
  dots). All doublets are preserved, but triplets and quadruplets disappear in the second case, as
  shown in panels $h=4,8,12,16,20$.}
\end{figure*}

\begin{figure*}[t]
\begin{center}
  \includegraphics[scale=0.53]{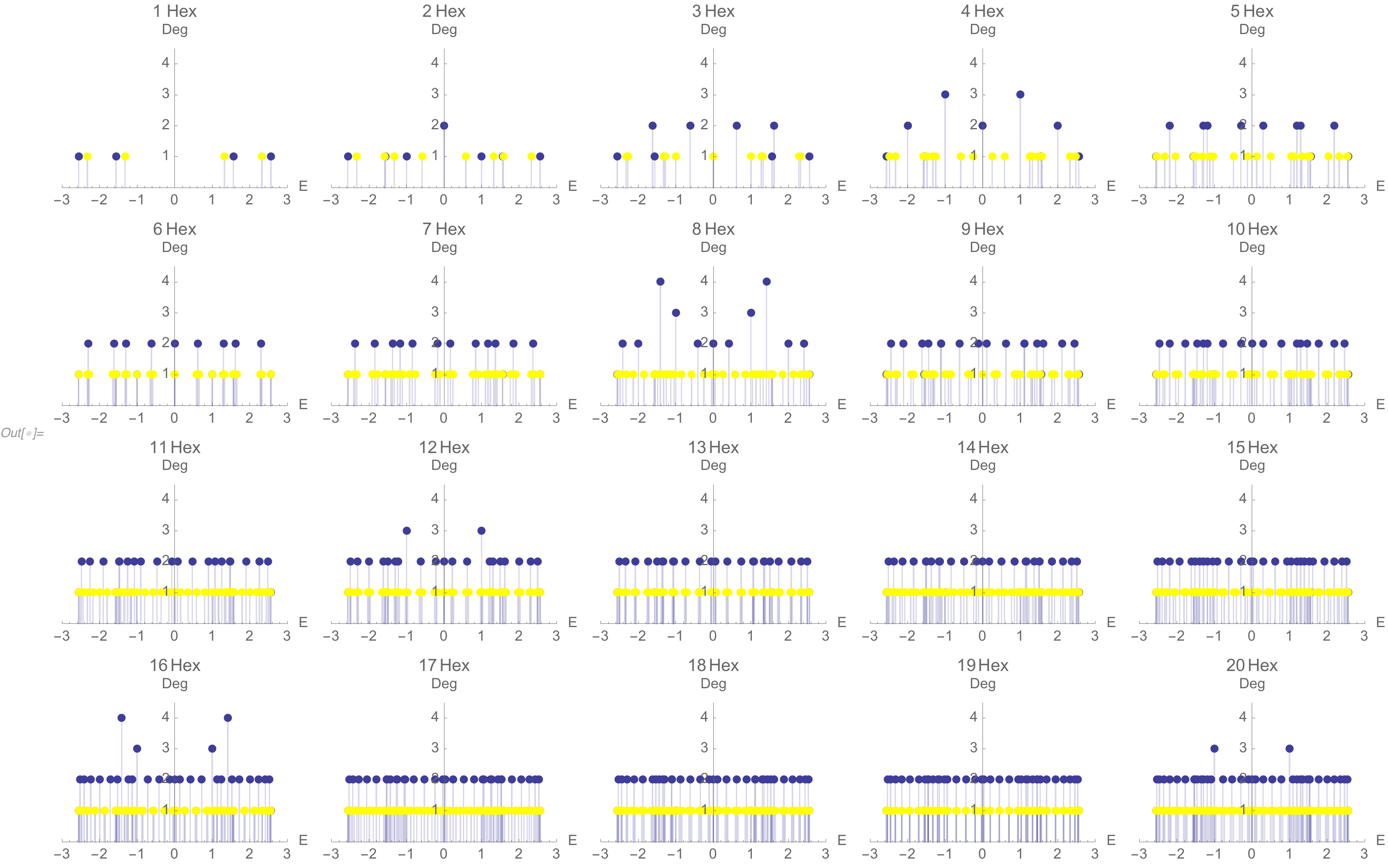}
\end{center}
\caption{\label{fig:s2} Energies without magnetic fields (blue dots) and with $\Phi_{B}=1$ (yellow
  dots). All doublets are broken, including all other degeneracies in triplets and quadruplets, as
  shown in panels $h=4,8,12,16,20$.}
\end{figure*}

\subsection{Symmetry breaking}

Various types of symmetry breaking reveal the nature of degeneracies in (\ref{s8}). Kramers
degeneracy and accidental triplets and quadruplets can be lifted by 1) the introduction of magnetic
fields, and 2) strain in the molecular structure, modifying the up-down couplings and the aspect
ratio of the hexagonal cells. Let us address first this mechanical deformation: The Hamiltonian $H$
in (\ref{s4}) suffers the modification $\v 1_{4\times 4} \mapsto \Delta \v 1_{4\times 4}$, where the
new parameter in terms of neighboring $\pi$ orbitals on sites $a,b$ is such that
\be
\Delta = \int dV  \pi_{a}(\v r)^*  \left[ \frac{p^2}{2m} + V_{ \small \rm Coulomb\ } \right]
\pi_{b}(\v r) \approx \Delta(0) \exp( |a-b|/\lambda),
\label{s11}
\ee
where $\lambda$ is the effective Bohr radius. So, varying the distance $|a-b|$ between up and down
chains modifies $\Delta$, and with such a replacement the Hamiltonian
\bea
H = \sigma_{+} K + \sigma_{-} K^{\dagger} + \frac{1}{2}\left( \Delta \v 1_{4\times4} + \sigma_3 \right) \tau_3
\label{s12}
\eea
has a modified spectrum
\bea
E_{k,s,t} = \frac{t + s \sqrt{ \Delta^2 + 16 \cos^2 (k/2) }}{2}.
\label{s13}
\eea
It is easy to see that the accidental degeneracy in triplets and quadruplets for the sequence
$h=4, 8, 12, 16...$ is lifted; this means that the building blocks of cyclopolyacene (benzene,
diallyl, anthracene and tetramethyl-naphtalene) are no longer isospectral, as the hexagons have lost
their $C_{v6}$ symmetry -- See fig. \ref{fig:s1}. However, doublets are still ruled by Kramers
degeneracy, so now we introduce a magnetic field.

When a magnetic flux $\Phi_{B}$ with field intensity $B$ is perpendicularly applied to a
cyclopolyacene ring (but parallel to the hexagonal planes!) time reversal symmetry is broken. In a
specific gauge, a phase factor enters the Hamiltonian with the modification of the kinetic term
$K \mapsto K(\Phi)$ and the eigenvalues of $K(\Phi)$ are obtained by the replacement
$1 + e^{ik} \mapsto 1 + e^{i (k+\beta)}$ yielding
\be
E_{k,s,t}(B) = \frac{t + s \sqrt{ 1 + 16 \cos^2 [ (k+ \beta)/2] }}{2}, \quad \text{with} \quad
\beta = \frac{e \Phi_{B}}{\hbar c}, \quad \Phi_{B} = B \times \mbox{Area}.
\label{s14}
\ee
This phase shift $\beta$ in the spectral formula breaks the old symmetry $k \mapsto 2\pi - k$ or
$q \mapsto h - q$ in (\ref{s8}), which reproduces as a consequence, the spectral version of the
Aharonov-Bohm effect, see fig. \ref{fig:s2}.

\subsection{Wave functions}

\begin{figure}[t]
  \begin{center}
    \includegraphics[scale=0.45]{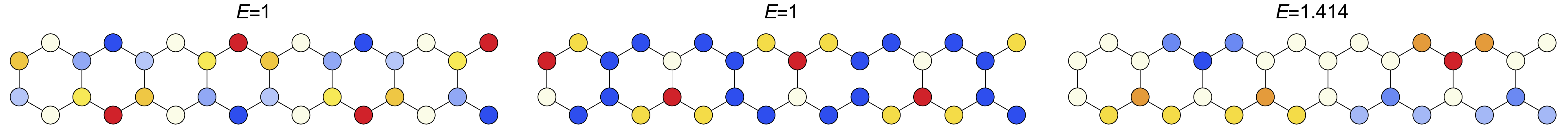}
  \end{center}
  \caption{\label{fig:sw1} Linear combinations of degenerate eigenstates of cyclopolyacene ($h=8$,
    $q=1$). In the case of triples $E=1$ disconnected hexagonal cells and a zigzag structure can be
    observed. In the case of quadruplets $E=\sqrt{2}$ a zigzag nodal structure.}
\end{figure}

Some waves supported by these structures have a peculiar nodal structure. This is so, because linear
combinations of degenerate waves may conspire to cancel the amplitude in various sites, without
destroying the periodicity of the array. It is mandatory that the waves are real,
i.e. $\Phi_{B}=0$. In the case $h=8$ and $q=1$, the disconnected segments are either chains of
hexagons resembling anthracene or chains of diallyl structures that build
tetramethyl-naphtalene. This depends on the choice of the two nodes on the polymer. To illustrate
this, we show in fig.~\ref{fig:sw1} such linear combinations of triplets and quadruplets,
respectively.

\section{Partial isospectrality of anthracene and tetramethyl-naphtalene}

Here we obtain the exact energies of anthracene and tetramethyl-naphtalene and show, with the help
of algebraic properties fulfilled by their Hamiltonians, that some of their levels overlap. This
will prove that cyclopolyacene has extraordinary degeneracies for $h=4,8,12,16,...$

The sketch of our proof is as follows: 1) Triplets emerge from benzene and diallyl segments, which
are in turn obtained from the antisymmetric functions of anthracene and tetramethyl-naphtalene. 2)
The Hamiltonian $Q$ of tetramethyl-naphtalene can be put in terms the Hamiltonian $H$ of anthracene;
from here and with some algebraic steps we show that 3) the corresponding spectra have reflection
properties around $E=0$ and addition properties connecting $H$ and $Q$; thus a doublet in $Q$ will
imply a doublet with the same energies in $H$, and because of spectral reflection, two other
doublets will appear at negative energies. Therefore two quadruplets arise in the full spectrum of
cyclopolyacene when $h=8, 16, 24,...$. In addition 4) the computation of $H^2$ and $Q^2$ in terms of
themselves will give access to analytic formulae for $\sigma(H)$ and $\sigma(Q)$.

\subsection{Exact solutions and algebraic properties}

According to the skeletal forms of anthracene (three hexagonal rings) and tetramethyl-naphtalene
(two hexagonal rings with additional leads to the left and the right), the antisymmetric functions
around the center break effectively the molecules into two benzene rings and two diallyl structures,
respectively. See fig.~\ref{fig:1.2} and \ref{fig:m5}. For unit nearest neighbor couplings, the
corresponding energies are
\be
E_{\mbox{\small C}_6} = -2,-1,-1,+1,+1,+2, \quad
E_{ 2 \mbox{\small C}_3} = -2,-1,0,0,+1,+2.
\label{s15}
\ee
So, at least these energies should be contained in the corresponding spectra. Now we proceed to
analyze the Hamiltonians $H$ and $Q$ (given by $14 \times 14$ matrices!). These are defined in
compact form with the help of a kinetic operator $\Pi$ and up-down couplings $D, \bar D$:
\bea
\Pi &=& \sum_{n=1}^{6}\left\{ | \mbox{up}, n \>\<  \mbox{up}, n{+}1 |
  + | \mbox{down}, n \>\<  \mbox{down}, n | \right\}+ \mbox{h.c.}, \nonumber \\[2mm]
D &=& \sum_ {n=1}^{4} | \mbox{up}, 2n{-}1 \>\<  \mbox{down}, 2n{-}1 |
+ \mbox{h.c.}, \quad \bar D = \v 1_{7\times 7} - D.
\label{s16}
\eea
When arranged in subspaces of even-odd sites and up-down sub chains, these matrices divide the full
Hamiltonian $H$ and $Q$ into four blocks of size $7\times 7$, as given by
\bea
H= \left( \begin{array}{cc}  \Pi &  D   \\  D   &  \Pi \end{array}   \right), \quad
Q= \left( \begin{array}{cc}  \Pi &  \bar D   \\  \bar D   &  \Pi    \end{array} \right).
\label{s17}
\eea
Simple unitary transformations turn these expressions into
\bea
\tilde H = U^{\dagger} H  U =  \left( \begin{array}{cc}  \Pi {+} D &  0   \\  0   &  \Pi {}- D \end{array}   \right), 
\quad U = \frac{1}{\sqrt{2}}  \left( \begin{array}{cr}  1 &  1   \\  1   &  -1  \end{array}  \right), \nonumber \\[2mm]
\tilde Q = V^{\dagger} Q  V =  \left( \begin{array}{cc}  \Pi {-} \tilde D &  0   \\  0   &  \Pi {+} \tilde D \end{array}   \right),
\quad V = \frac{1}{\sqrt{2}}  \left( \begin{array}{cr}  1 &  -1   \\  1   &  1  \end{array}  \right), \nonumber\\
\label{s18}
\eea
which proves immediately the important relation
\bea
\tilde H = \tilde Q + \left(\begin{array}{cc}  \v 1_{7 \times 7} &  0   \\  0   &  -\v 1_{7 \times 7}  \end{array}  \right). 
\label{s19}
\eea
The spectra of $H, \tilde H$ coincide, as well as those of $Q, \tilde Q$. The relation above further
implies that $\sigma(H)$ and $\sigma(Q) \pm 1$ are related. Now we prove a specularity property of
the two spectra around $E=0$ with an operator $P$ written in two different representations as

\be
P= \sum_{n=1}^{7} (-1)^n | n \>\< n | =
\left(
  \begin{array}{ccccccc}
    -1 &&&&&&\\ &-1&&&&&\\ &&-1&&&& \\ &&&-1&& &\\ &&&&+1&& \\ &&&&&+1& \\ &&&&&&+1
  \end{array} \right). 
\label{s20}
\ee
The following useful relations hold
\be
P^2 =1 = P P^{\dagger},\quad \left\{P, \Pi \right\} = 0, \quad [P, D] = [P, \tilde D]=0.
\label{s21}
\ee
As a result, the blocks conforming $\tilde H$ and $\tilde Q$ have the isospectral relations
\bea
P (\Pi + D) P^{\dagger} = -\Pi + D = -(\Pi + \tilde D) + \v 1_{7 \times 7}, \nonumber\\[2mm]
P (\Pi - D) P^{\dagger} = -\Pi - D = -(\Pi - \tilde D) - \v 1_{7 \times 7}.
\label{s22}
\eea
This shows that $H$ and $Q$ have specularity around $E=0$, because $P$ maps unitarily upper blocks
of $\tilde H$ and $\tilde Q$ into their negative lower blocks and vice versa. Also, the upper block
of $\tilde H$ is mapped into the lower block of $\tilde Q$ plus 1. On the other hand, the individual
blocks $\Pi \pm D$ have no degeneracy, so the doublets we are looking for must be levels shared in
common by $\Pi + D$ and $\Pi - D$. To see this, we make use of the following algebraic relations
\be
\{\Pi , D \} = \Pi, \quad \{\Pi, \tilde D \} = \Pi, \quad
\Pi^2 = (\Pi {+} D)^2 {-} (\Pi {+} D) = (\Pi {+} \tilde D)^2 {-} (\Pi {+} \tilde D)
\label{s23}
\ee
and therefore the spectrum of the block $\Pi + D$ can be put in terms of $\Pi^2$, which is very easy
to compute
\be
\Pi^2 = \left( \begin{array}{cc} h_1 & \\ & h_2 \end{array} \right), \: \:
h_1 = \left( \begin{array}{cccc} 1&1&0&0 \\ 1&2&1&0 \\ 0&1&2&1 \\ 0&0&1&1  \end{array} \right), \:\:
h_2 = \left( \begin{array}{ccc} 2&1&0 \\ 1&2&1 \\ 0&1&2  \end{array} \right).
\label{s24}
\ee
This shows that $\Pi^2$ is equivalent to two disconnected chains of length 3 and 4, always solvable
by radicals, with the resulting energy levels:
\be
\sigma(h_1) = 0, 2-\sqrt{2},+ 2, 2 + \sqrt{2}, \quad \sigma(h_2) = 2-\sqrt{2}, +2, 2 + \sqrt{2}.
\label{s25}
\ee
Finally, we put the spectrum $E_{+}$ of $\Pi + D$ in terms of $\Pi^2$ by solving
$E(\Pi^2) = x^2 - x$ for $x$ according to what we found in (\ref{s23}), and the spectrum $E_{-}$ of
$\Pi-D$ is obtained by solving $E(\Pi^2) = x^2 + x$ also for $x$:
\be
E_{+} = \frac{1 \pm \sqrt{1+4 E(\Pi^2)}}{2} \neq 0, \quad
E_{-} = \frac{-1 \pm \sqrt{1+4 E(\Pi^2)}}{2} \neq 0,
\label{s26}
\ee
leading to
\be
E_{+} = \frac{1 \pm \sqrt{9 \pm 4 \sqrt{2}}}{2}, \frac{1 \pm 3}{2}, \frac{1 + 1}{2} 
= -\sqrt{2}, -1, 1-\sqrt{2}, +1,+\sqrt{2},+2,1+\sqrt{2}
\label{s27}
\ee
and similarly
\be
E_{-} = -1-\sqrt{2},-2,-\sqrt{2}, -1, \sqrt{2}-1, +1, +\sqrt{2}. 
\label{s28}
\ee
The levels in common between $E_{+}$ and $E_{-}$ (upper and lower blocks) are as expected,
$E=\pm \sqrt{2}$ and $E=\pm 1$; the first doublets are shared by $H$ and $Q$, while $\pm1$ are
doublets for $H$ but singlets for $Q$. For completeness, we report the levels of $Q$ in agreement
with the second relation in (\ref{s15})
\be
E_{Q} = -1 - \sqrt{2}, 1 + \sqrt{2}, -2,+ 2, -\sqrt{2}, -\sqrt{2}, +\sqrt{2}, +\sqrt{2}, -1, +1, 1 -
\sqrt{2}, -1 + \sqrt{2}, 0, 0.
\label{s29}
\ee
It is amusing to confirm the derived properties $E \mapsto -1 + E$, $E \mapsto - E$ for some
elements in the sets (\ref{s27}) and (\ref{s28}). We conclude by noting that this procedure can be
generalized to open chains of an arbitrary number of hexagons, as the algebraic relations
(\ref{s19})-(\ref{s24}) also hold with the appropriate dimensional extensions.

\section{Exact energy levels and wave functions for M\"obius cycloacene}

The Hamiltonian of a M\"obius cycloacene can be represented in a simple manner, by recognizing the
topological equivalence between the compound and a circular nearest-neighbor graph with additional
couplings connecting odd sites with their antipodes. In such form, the Hamiltonian reads
\be
H = \left( \begin{array}{cccccccccccccc}  0 & 1 & 0 & 0 & \cdots &0&0& c & 0 & 0 & 0 &  \cdots & 0 & 1 \\ 
1 & 0 & 1 & 0 & \cdots & 0& 0 & 0 & 0 & 0 & 0 & \cdots & 0 & 0 \\ 
0& 1 & 0 & 1 & \cdots & 0 & 0 & 0&0&c&0 & \cdots &0&0 \\
0&0&1&0 &\cdots &0&0&0&0&0&0&\cdots&0&0 \\
 \vdots & & & & \ddots  & & & \vdots & & & & \ddots  & & \vdots \\
0&0&0&0& \cdots & 0& 1 & 0&0&0&0 &\cdots & c & 0 \\
0&0&0&0& \cdots & 1& 0 & 1&0&0&0 &\cdots & 0 & 0  \\
c & 0 & 0 & 0 & \cdots & 0&1&0&1&0&0&\cdots&0&0 \\  
0&0&0&0&\cdots&0&0&1&0&1&0 &\cdots & 0&0    \\ 
0&0&c&0& \cdots &0&0&0&1&0&1& \cdots &0& 0 \\ 
0&0&0&0& \cdots &0&0&0&0&1&0& \cdots &0&0 \\ 
\vdots &&&&\ddots&&&&&&&\ddots&& \\
0&0&0&0 &\cdots& c&0 &0&0&0&0 &\cdots &0 &1 \\
1&0&0&0& \cdots &0&0&0&0&0&0&\cdots& 1& 0
\end{array} \right),
\label{s30}
\ee
where $c$ is the vertical coupling of hexagons. In a regular geometry one has $c=1$. Now we try the
following basis
\bea
| q, 1 ) &&= \frac{1}{\sqrt{2h}}\sum_{n=1}^{4h} \left( \frac{1-(-1)^n}{2} \right) \exp[iq(n+1)/2]|n\>,
\nonumber \\
| q, 2 ) &&= \frac{1}{\sqrt{2h}}\sum_{n=1}^{4h} \left( \frac{1+(-1)^n}{2} \right) \exp[iqn/2]|n\>,
\label{s31}
\eea
where $|n\>$ is the canonical basis and $h$ is the number of hexagons. These vectors correspond to
even and odd sites, respectively. Upon translations of two units, they satisfy
$T^2 |q,a) = e^{iq} |q,a)$. By using the double periodicity of the
system (note that $2h$ sites are required to return to the original atom) one finds
$(e^{iq})^{2h}=1$ or $ q = n \pi / h$ where $n=1,...,2h$.

The application of $H$ to $|q,a)$, $a=1,2$ leads to a closed result within the basis. Using
(\ref{s30}) and (\ref{s31}), one has

\bea
H |q,1) &&= c e^{iqh} |q,1) + (1+e^{iq}) |q,2) \nonumber \\[2mm]
H |q,2) &&= (1+e^{-iq}) |q,1).
\label{s32}
\eea
We note here that, by construction of the basis, the coefficient $e^{iqh}=(-1)^n$ is
real. Therefore, we have the following Hermitian block ready for diagonalization:
\bea
H^{(q)} = \left( \begin{array}{cc} c e^{iqh}  & 1 + e^{iq} \\  1 + e^{-iq} & 0 \end{array} \right).
\label{s33}
\eea
The energy formula for general vertical coupling $c$ reads
\bea
E_{n}^{\pm} = \frac{ (-1)^{n} c \pm \sqrt{c^2 + 16 \cos^2 \left( \frac{n \pi}{2 h} \right)  } }{2}.
\label{s34}
\eea
Eigenvectors can be easily found for each $2\times 2$ block:
\bea
|E, 1 \> &=&  \cos(\theta/2) | q,1) + e^{i\phi} \sin(\theta/2) | q,2), \nonumber\\[2mm]
|E, 2 \> &=& -\sin(\theta/2) | q,1) + e^{i\phi} \cos(\theta/2) | q,2),
\label{s35}
\eea 
with $\cos \theta = c (-1)^n / (2+2\cos q )$, and $\tan \phi = - \sin q / (1 + \cos q)$.

\end{document}